\documentclass{article}
\usepackage{geometry}
\usepackage{bm,epsfig,ifthen}
\usepackage{amsfonts,amssymb}
\usepackage[english]{babel}
\usepackage[latin1]{inputenc}
\usepackage{graphicx}
\usepackage{amssymb,amsmath}
\usepackage{pstricks}
\def\rg{R_{\rm g}}
\def\rt{R_{\rm t}}
\title{Mechanical and statistical study of the laminar hole formation in transitional plane Couette flow}
\author{Joran Rolland\footnote{LadHyX, UMR 7646 CNRS, Palaiseau 91128, France \& INLN, UMR 7335 CNRS, UNSA, 1361 route des lucioles, 06560 Valbonne, France} \footnote{rolland@ladhyx.polytechnique.fr}}
\date{Submitted to Eur. Phys. J B: 28/05/2014, accepted: 03/02/2015.}

\begin{document}
\maketitle
\begin{abstract}
This article is concerned with the numerical study and modelling of two aspects the formation of laminar holes in transitional
turbulence of plane Couette flow (PCF). On the one hand, we consider quenches: sudden decreases of the Reynolds number $R$ which force the formation of holes.
The Reynolds number is decreased from featureless turbulence to the range of existence of the oblique laminar-turbulent bands $[R_{\rm g}; R_{\rm t}]$.
The successive stages of the quench are studied by means of visualisations
and measurements of kinetic energy and turbulent fraction. The behaviour of the
kinetic energy is explained using a kinetic energy budget: it shows that viscosity causes quasi modal decay until lift-up equals it and creates a new balance. Moreover, the budget confirms that the physical mechanisms at play are independent of the way the quench is performed. On the other hand we consider the natural formation of laminar
holes in the bands, near $R_{\rm g}$. The Direct Numerical simulations (DNS) show that holes in the turbulent bands provide a
mechanism for the fragmented bands regime and orientation fluctuations near $R_{\rm g}$. Moreover the analysis of the fluctuations
of kinetic energy toward low values demonstrates that the disappearance of turbulence in the bands can be described within the
framework of Large Deviations. A Large Deviation function is extracted from the Probability Density Function of the kinetic energy
 \end{abstract}
 \begin{flushleft}Keywords: Shear turbulence -- Transition to turbulence -- Nonequilibrium Statistical physics and Large Deviations, 
\end{flushleft}
\begin{flushleft}
\end{flushleft}

\section{Introduction}

\begin{figure}
\centerline{\textbf{(a)}\includegraphics[width=5cm,clip]{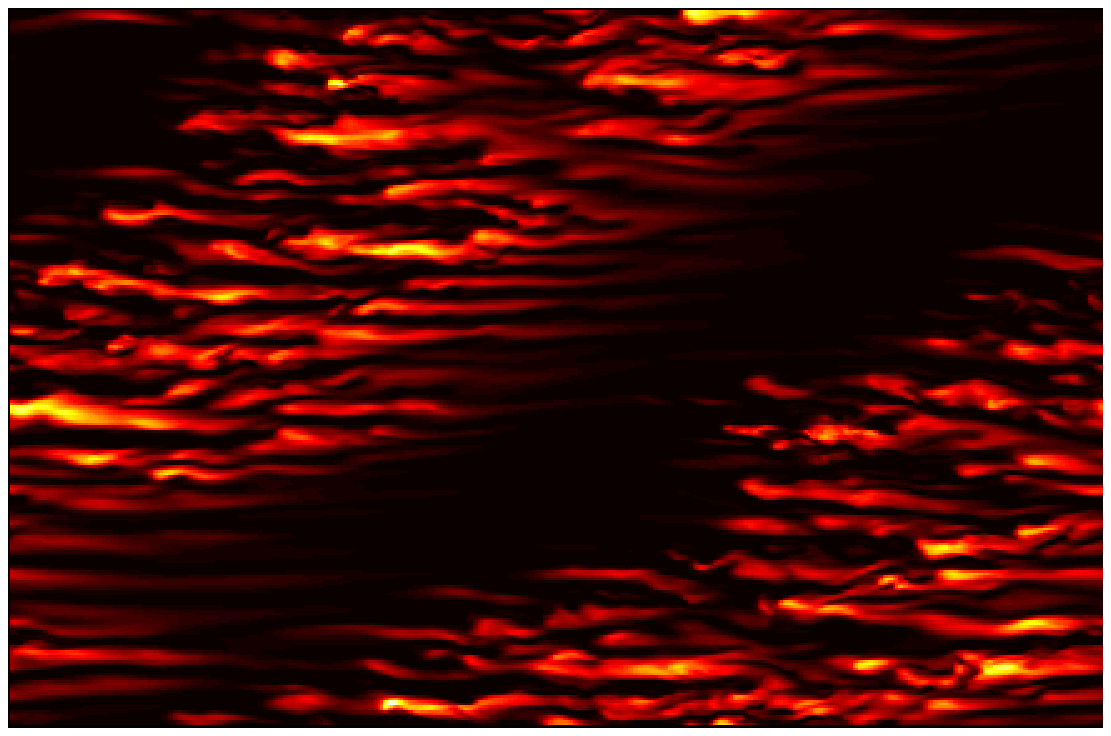}\textbf{(b)}\includegraphics[width=8cm]{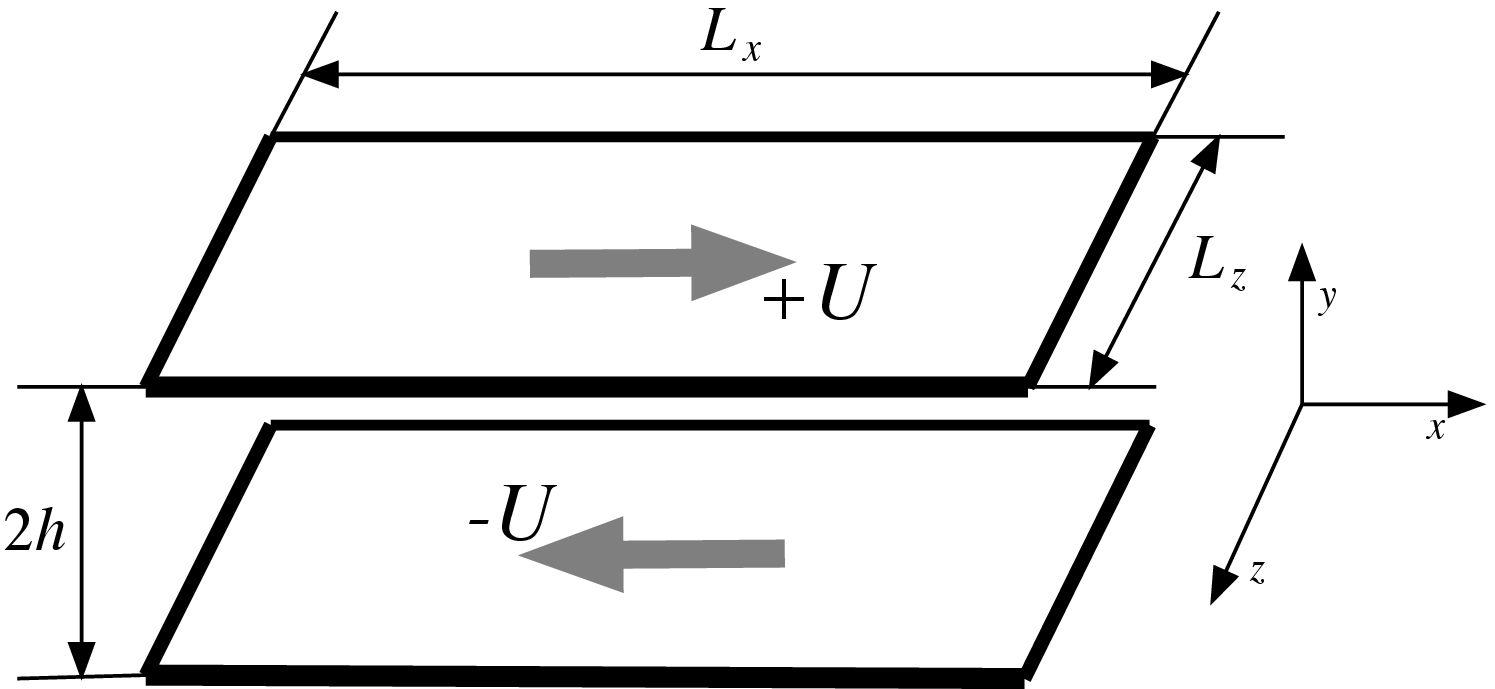}}
\caption{(a): Colour levels of the norm of the departure to the baseflow, result of a DNS in a periodic domain of size $L_x\times L_z=110\times72$, at $R=350$ (See Tab.~\ref{tab1} $\sharp 3$ for settings). (b): Sketch of PCF, introducing the domain size and velocity of the plates.}
\label{figint}
\end{figure}

Using  approaches from mechanics and statistical physics, this article examines the formation of laminar holes in featureless turbulence
or in the oblique band regime of transitional plane Couette flow (Fig.~\ref{figint} (a)). Like the other shear flows, PCF, the flow between
two parallel moving plates (Fig.~\ref{figint} (b)), is always linearly stable. As a consequence, turbulence can only be triggered by finite amplitude disturbances.
The transition is controlled by the Reynolds number of the flow. Above a Reynolds number $\rt  \simeq 415$ the turbulence is
said to be featureless and invades the whole domain \cite{prls}. If the Reynolds number is decreased below $\rt$ and kept above $\rg\simeq325$,
the flow displays the intriguing oblique bands regime (Fig.~\ref{figint} (a)) \cite{RM10_1,BT11,phD}. The coexistence of laminar and turbulent flow is spatially regular and statistically steady. Around $\rg$, the flow displays complex
dynamics of local relaminarisations \cite{sch}, fragmented bands \cite{phD,prigent02}, spot splitting \cite{shi} and growth or retraction of spots \cite{prls,Yms,ispspot}.

Understanding the formation of laminar holes
(or laminar troughs), that is region of very low kinetic energy in otherwise turbulent flow,
is a key to explain the laminar turbulent coexistence. For that matter,
two types of laminar troughs can be distinguished. On the one hand, one finds the troughs appearing in featureless turbulent flow,
after a sudden decrease of the Reynolds number termed quench \cite{phD,bddm}. These holes lead to the statistically steady oblique bands regime.
On the other hand, laminar troughs can appear naturally inside the bands, and may lead to the collapse of turbulence \cite{sch,shi,M11,io}.
The first type of laminar hole formation is easier to capture, since it is bound to happen in a finite time whenever the Reynolds number is decreased below $\rt$.
Assuming that the mechanisms are very similar in both cases, the study of quenches can bring insight to the understanding of the collapse of turbulence and \emph{vice versa}.
These two points of view provide an occasion to test two different approaches of modelling.

   At the scale of the band, or that of the coherent structure, energy budgets \cite{schhu} can be derived from Navier--Stokes equations.
The comparison of the budget to the data extracted from DNS can give simple arguments to explain the behaviour of the flow.
A mechanism can be proposed after the identification of the weight of dissipation (viscosity) and  production in the variation of the kinetic energy.
Production is caused by a central process in sustainment of turbulence: lift-up, that is, the extraction of energy from the mean flow by streamwise vortices \cite{W} (See \cite{brandt_palm} for a recent review in Newtonnian and non-Newtonnian flows).

  An new point of view has been proposed to explain
the super exponential dependence of the life time of turbulence in pipe flow: extreme fluctuations of the velocity fields leading to the collapse of turbulence \cite{gold}.
The study of such rare events is often undertaken in the framework of Large Deviations \cite{ht}.
This point of view completes the classical approaches taken
to understand the disappearance of turbulence near $R_{\rm g}$: dynamical systems theory applied to the chaotic evolution of the self sustained process of
turbulence \cite{fe04,W} or application of the spatiotemporal intermittency approach of turbulent convection \cite{shi,Cprl,cm,B}.
Indeed, it allows one to compute escape probabilities or lifetimes from the properties of the phase space of Navier--Stokes equations.
In that spirit, long exponential tails of the Probability Density Function (PDF) of the kinetic energy have been measured in modelling
of PCF \cite{M11}, and their asymmetry has been quantified \cite{io}. Using an analogy with few degrees of freedom models, one can argue that this asymmetry is the precursor of a transition. However, several questions remain open. Does the collapse of turbulence
in wall bounded flows fit in the general framework of Large Deviations ? Can all the concepts and methods associated to Large Deviations
be used to model and predict the transition ? The study formation of holes near $R_{\rm g}$ is adapted to answer these  questions.

   The investigation of the properties of hole formation from these two points of views structures the organisation of the paper.
The DNS procedures and the quantities used to investigate the flow are described first (\S~\ref{proc}). The quench numerical
experiments are then presented (\S~\ref{Q}), seen at the scale of the band (\S~\ref{Q1}) and at the scale of the velocity streaks and the streamwise vortices (\S~\ref{locm}).
We then suggest and validate a model describing the behaviour of the flow during the quench (\S~\ref{M}). We eventually move to the
study of laminar holes appearing in bands near $\rg$ (\S~\ref{H}).These results are discussed together in section~\ref{D}.

\section{Procedure \label{proc}}

\subsection{Notations and numerical procedure\label{proc1}}

  Plane Couette flow is the flow between two parallel plates, at fixed positions $y=\pm h$. The two plates move in opposite directions at velocities $\pm U \mathbf{e}_x$ (Fig.~\ref{figint} (b)).
Making the system dimensionless simplifies comparisons with other flows, particularly with laboratory experiments: it largely reduces the number of control parameters.
Moreover, provided the non-dimensionalisation is based on the proper physical criteria,
the range of control parameter at which the transition occur is the same in all wall bounded flows \cite{BT11}. Lengths are in units of $h$, velocities in units of $U$
(the velocity of the top plate) and times in units of $h/U$. The dimensionless velocity field is $y \mathbf{e}_x+\mathbf{v}$ with $\mathbf{v}$ the departure from the laminar base flow.
The Reynolds number $R=hU/\nu$, with $\nu$ the kinematic viscosity, is the control parameter of the system.
Together with the sizes $L_x$ and $L_z$, it controls the behaviour of the turbulent flow.

  The spatial discretisation of the velocity field $\textbf{v}$ and its integration in time is based on the code {\sc ChannelFlow} written by John Gibson \cite{gibs}.
Fourier decomposition is used in the streamwise ($x$) and spanwise ($z$) directions, with $N_{x,z}$ dealiased modes. Meanwhile, $N_y$ Chebychev polynomials
are used in the wall-normal direction ($y$). In the plane, boundary conditions are periodic, while no slip conditions are used at the walls.
An in plane resolution of $N_{x,z}/L_{x,z}=4$ and a wall-normal resolution of $N_y=27$ was found to be sufficient to perform a DNS of the flow at these Reynolds numbers \cite{MR10,dsc10}.
The code was used with a reduced resolution $N_{x,z}/L_{x,z}=8/3$, $N_y=15$ as a low order model of the flow. It is used when long simulations or repetitive statistics  have to be performed (cumulated durations of several $100000h/U$) \cite{RM10_1,RM10_2}. The oblique bands regime was found to be well rendered at this resolution.
The reduction of resolution affects mostly the transition thresholds. They are reduced from $[\rg;\rt]\simeq[325; 415]$, at $N_y=27$, to  $[\rg;\rt]\simeq[275; 355]$, at $N_y=15$ \cite{MR10}. For each domain size used in this article, the resolutions used and the corresponding thresholds are summarised in table~\ref{tab1}.

\begin{table}
\begin{center}\begin{tabular}{|c|c|c|c|}
\hline
$\sharp$&$L_x\times L_z$ &$N_{x,z}$, $N_y$&$[\rg;\rt]$ \\ \hline
1&$110\times 32$ & 4, 27&$[325; 415]$\\ \hline
2&$56\times 48$ & 4, 27& $[325; 415]$ \\ \hline
3&$110 \times 72$ & 4, 27 & $[325; 415]$\\ \hline
4&$110\times 64$ & 8/3, 15 &$[275; 355]$\\ \hline
5&$220\times 48$ & 8/3, 15&$[275; 355]$ \\ \hline
6&$440 \times 48$ & 8/3, 15&$[275; 355]$ \\ \hline
7&$660 \times 48$ & 8/3, 15&$[275; 355]$ \\ \hline

\end{tabular}
\end{center}
\caption{Table summarising the settings number ($\sharp$), the size used and the corresponding resolutions and thresholds.}
\label{tab1}
\end{table}

\subsection{Measurements\label{proc2}}

In order to monitor the state of the flow, we compute the spatially averaged kinetic energy, $e=\frac{1}{2}\int {\rm d}x{\rm d}y{\rm d}z\parallel\textbf{v}\parallel^2$, from the square of the departure from the laminar plane Couette flow.
Besides, we compute the turbulent fraction $f$ (sometimes called the intermittency factor),
using a method introduced in a earlier studies of the bands \cite{RM10_1,ispspot,PM}. This method is based on the identification the turbulent regions of the flow.
The domain is divided in small boxes of size $l_x\times l_y\times l_z=2\times 1\times 2$, in which $\parallel\textbf{v}\parallel^2$ is locally averaged.
The energy in each cell is compared to a threshold $c=0.0125$ : if the local average of $\parallel\textbf{v}\parallel^2$ is above $c$, the cell is turbulent, otherwise, it is laminar.
This threshold has been chosen in a former study \cite{RM10_1}.
The turbulent fraction is then defined as the fraction of cells which are turbulent, \emph{i.e.} the relative volume occupied by turbulence in the flow.

A quantity indicating whether the turbulence is organised in bands and which orientation do the bands take is necessary.
For that matter, we take advantage of the sinusoidal modulation of turbulence of the band \cite{RM10_1,BT11,prigent02}.
Once the fundamental mode of the bands $(k_x,\pm k_z)$ is identified, two similar strategies can be followed to define a so-called order parameter.
The first one is adapted if the size of the system ($L_{x,z}$) is large relatively to the wavelengths of the bands $(2\pi/k_{x,z})$.
One can then compute two Hilbert transforms of the signal measured
around modes $(k_x,\pm k_z)$ \cite{phD,prigent02}. In numerical studies, the signal is the velocity field, while in the experimental studies, the signal is the light intensity. This yields two complex functions $a_+(x,z,t)$ and $a_-(x,z,t)$ which vary slowly in space relatively to the wavelength of the band.
 Their moduli give the amplitude of the modulation of the respective $\pm$ orientation at position $(x,z)$ at time $t$ while their phase give the relative shift of each patch of band.

 The second strategy is adapted if the size of the system is comparable to the wavelengths of the bands. Following Tuckerman \& Barkley \cite{BT11},
we proposed to use the Fourier transform of the $x$ component of the velocity field to define the order parameters $a_\pm=m_\pm(t)e^{\imath\phi_\pm(t)}$  by computing \cite{RM10_1}:
\begin{equation}
  m_\pm^2=\frac{1}{2}\int_{y=-1}^{y=1}|\hat{u}_x|^2(k_x,y,\pm k_z)\,{\rm d}y\,,\,
  \phi_\pm=\arg\left(\hat{u}_x(k_x,0,\pm k_z\right)\,.
\end{equation}
The phases $\phi_\pm$ of the order parameters give the relative position of the respective orientation of the band in the domain.
Meanwhile, the moduli of the order parameters give the amplitude of the modulation of the respective orientation. In this article, we follow this approach.
Since we do not study the behaviour of the phases, we only use $m_\pm$, termed the order parameters by an abuse of language.

The order parameters are the ideal quantities to study the behaviour of the band and its appearance from featureless turbulence at $R_{\rm t}$.
Unlike the turbulent fraction, they capture the spatial organisation of turbulence. This is particularly interesting when the two orientations of the band are competing.
The presence of the system in one state or another can be easily determined by the ratio $m_+/m_-$: if it is large, the orientation $+$ is observed, if it is small,
the orientation $-$ is observed. If it is of order $1$, a mix of both orientations, a regular lozenge \cite{ispspot} or a more defective coexistence of laminar and turbulent flow \cite{RM10_2},
is observed. The order parameters may not be as efficient as the turbulent fraction to characterise the change of regime at $R_{\rm g}$ \cite{shi}.
However, they still yield a precise information as to the organisation of the flow in the so-called fragmented bands regime \cite{phD}.

\section{Quenches \label{Q}}

\subsection{Principle}

  The first and most simple way to obtain and study laminar holes appearing in turbulent flows is to perform so-called quench experiments \cite{bddm}.
  A featureless turbulent flow, at $R_0>\rt$ ($R_0=500$ at $N_y=27$) is taken as an initial condition.
  It is obtained by integrating in time a random velocity field  until it reaches statistically steady wall-turbulence. At the initial instant of the quench experiment,
  the Reynolds number is suddenly decreased to the so-called arrival Reynolds number $R_1\ge \rg$. Note that $R_1$ can be above $\rt$. The simulation is performed at this constant value of the Reynolds number.

  Decorrelated featureless turbulent initial conditions are obtained by integrating in time a featureless turbulent velocity field over a duration of $300h/U$.
  This is useful for the study of the effect of the initial local organisation on the spatially averaged behaviour of the flow.

  Practically speaking, the quenches correspond to two possible situations. If the Reynolds number is decreased, and the velocity fields are unchanged,
they correspond to an increase by $R_0/R_1=(Uh/\nu_0)/(Uh/\nu_1)=\nu_1/\nu_0$ of the kinematic viscosity $\nu$. Meanwhile, if the velocity is rescaled by $R_0/R_1=(U_0h/\nu)/(U_1h/\nu)$
after the decrease of $R$, this corresponds to a decrease of the velocity of the plates by $R_1/R_0=U_1/U_0$. The first approach is studied extensively in this article.
The second is used in several cases in order to show that both approaches are equivalent.
Although the decrease of viscosity can only be achieved in experiments where a cooling system controls the temperature and therefore the viscosity,
it is the most systematic one numerically speaking.

  Both implementations depart slightly from a laboratory experiment. Indeed, in a laboratory experiment, the viscosity or the velocity profile have to adapt to
the new imposed boundary conditions. Both adaptations occur in a time scale respectively proportional to the inverse of the temperature diffusion coefficient
or the Reynolds number. Since {\sc ChannelFlow} solves the equations for the departure to the laminar plane Couette flow, this raises other implementation questions
such as using an actual discontinuity at $t=0$, causing numerical instability, or taking into account the inertia of an experimental apparatus. By testing both
the reduction of $U$ and the augmentation of $\nu$, we will argue in section~\ref{M} that our procedure captures the main physical mechanisms.

In that matter, Poiseuille
flows are probably the most adapted to the simulation of realistic quenches.
Indeed, the Reynolds number is decreased through a decrease of the pressure gradient or the flow rate, which adjust through the whole flow at the speed of sound.
The corresponding duration is much smaller than the diffusion time.

\begin{figure}\centerline{\textbf{(a)}\includegraphics[width=4cm,clip]{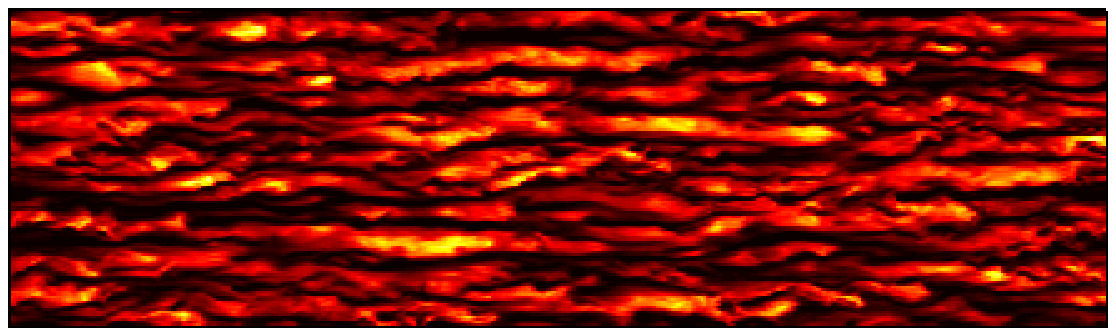}\hspace{1mm}\textbf{(b)  }\includegraphics[width=4cm,clip]{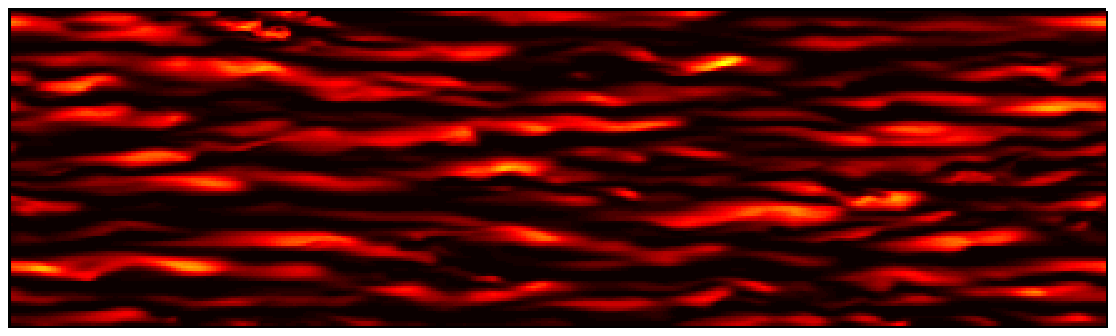}\hspace{1mm}\textbf{(c)}
\includegraphics[width=4cm,clip]{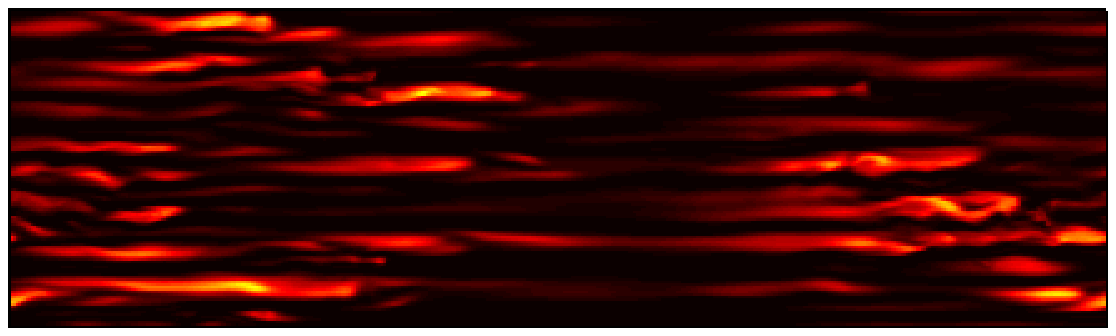}\hspace{1mm}\textbf{(d)}
\includegraphics[width=4cm,clip]{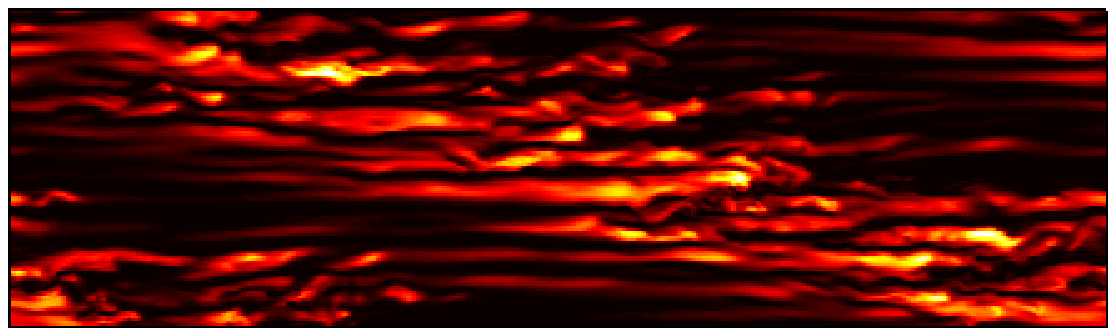}}
\caption{Colour levels of the norm of the departure to the laminar baseflow $\parallel\textbf{v}\parallel^2$ during a quench experiment,
in a domain of size $L_x\times L_z=110\times 32$, $R=370$. (a): $T=0$ , (b): $T=50$, (c): $T=150$, (d): $T=500$.}
\label{figim}
\end{figure}

\subsection{Visualisations and global measurements\label{Q1}}

\subsubsection{The quench}

\begin{figure}\centerline{\includegraphics[width=6cm,clip]{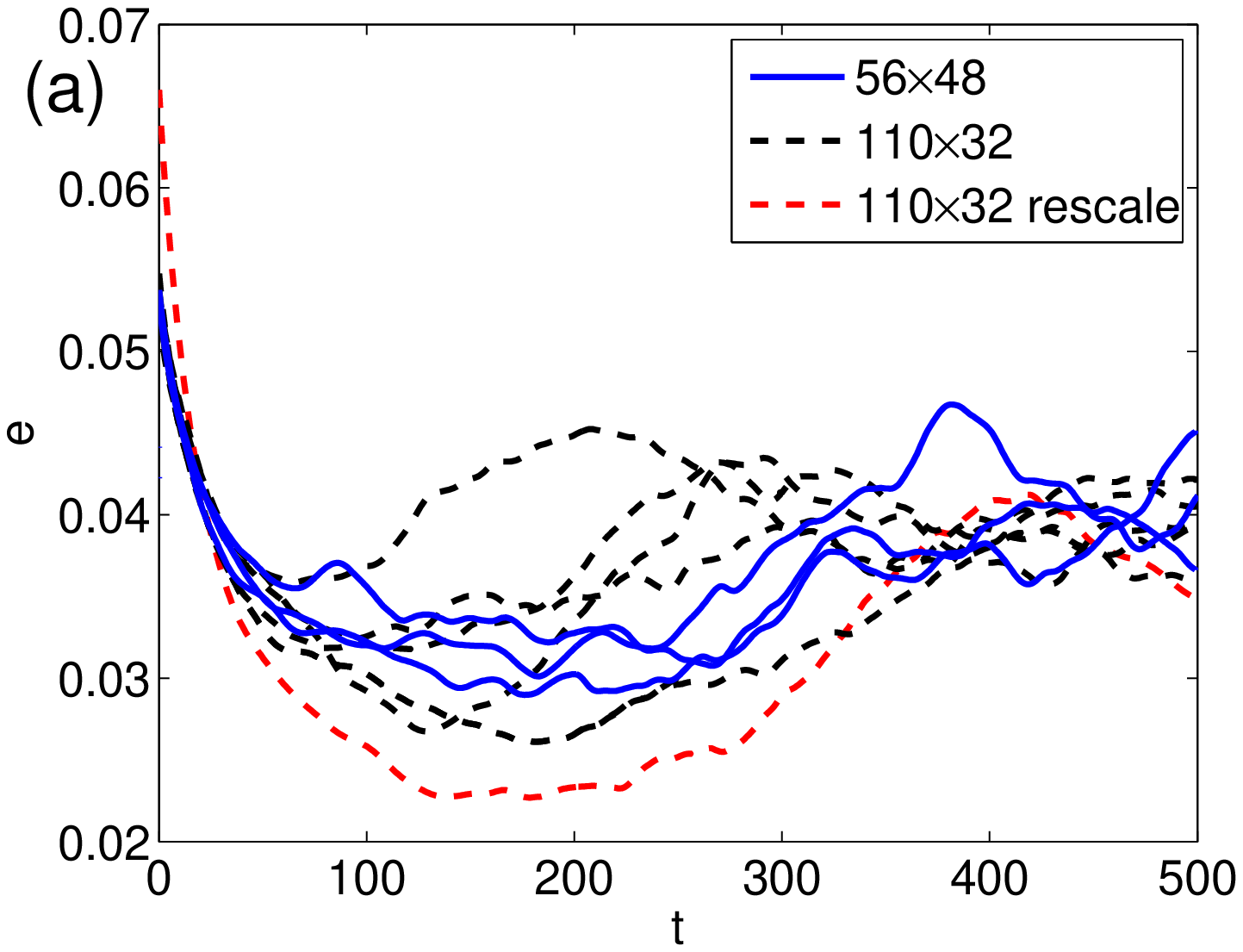}
\includegraphics[width=6cm,clip]{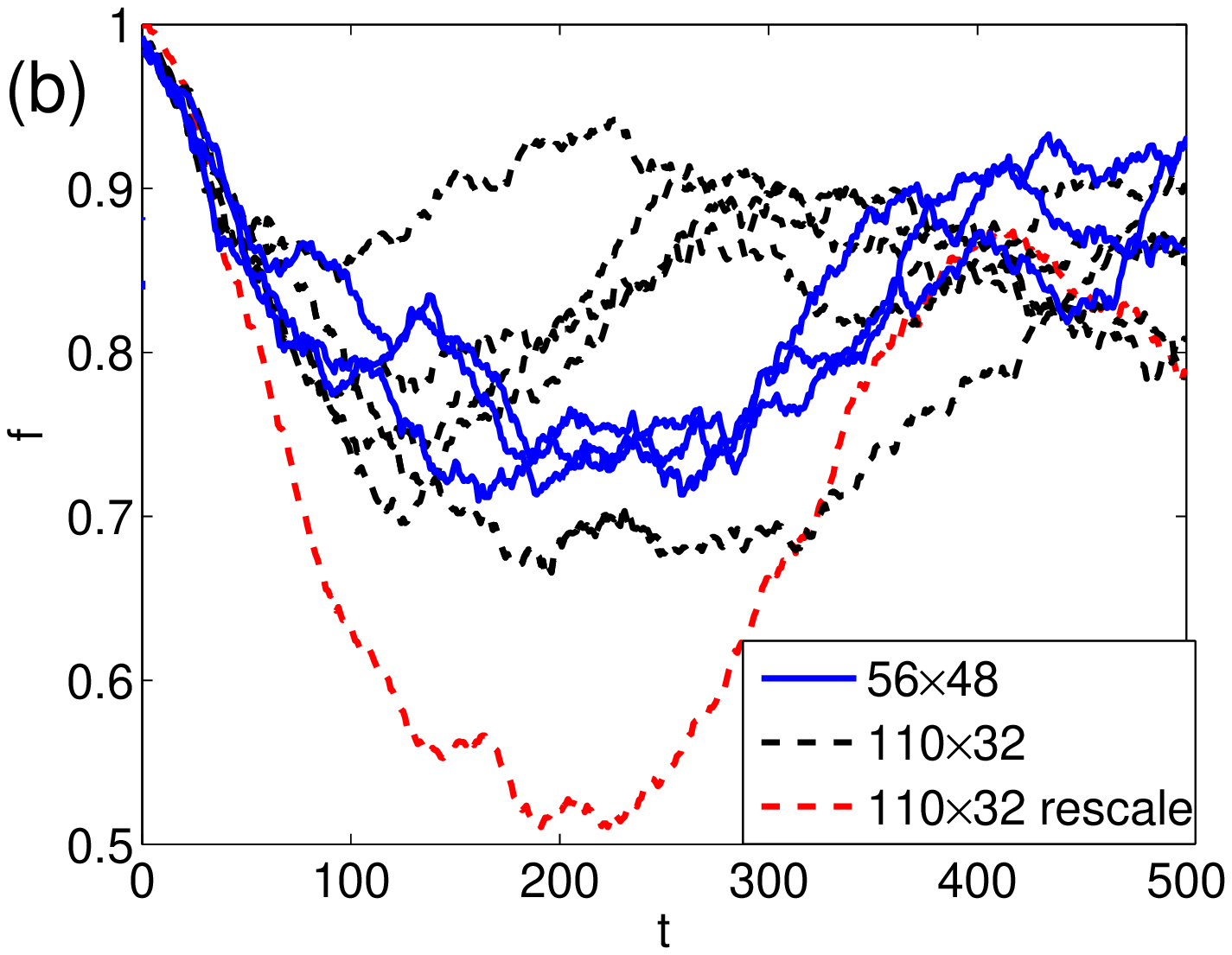}}
\caption{Times series for quenches ($R_0=500$,
$R_1=370$), in two domains of size $L_x\times L_z=56\times 48$ (full line) and $L_x\times L_z=110\times 32$ (dashed line), including a simulation with initial velocity rescaled by $R_0/R_1$, red). (a) : average kinetic energy $e$, (b) : turbulent fraction $f$.} \label{trempe3_bis}
\end{figure}

  The quench is first studied for a given arrival Reynolds number $R_1=370$ in two domains of sizes $L_x\times L_z=110\times 32$
  (with and without rescaling $\textbf{v}$ by $R_0/R_1$) and $L_x\times L_z=56\times 48$ (Tab.~\ref{tab1} $\sharp 1$ and $\sharp 2$). The first domain accommodates one wavelength of the band,
  of fundamental wavenumbers $k_x=2\pi/110, k_z=2\pi/32$. The second one only displays an intermittent laminar-turbulent coexistence \cite{PM}.
  In order to test the effect of the initial organisation of the flow on the quench, five decorrelated uniformly turbulent initial conditions are used for both sizes.
  We consider the time series of kinetic energy (Fig.~\ref{trempe3_bis}(a)) and turbulent fraction (Fig.~\ref{trempe3_bis}(b)), and compare them to visualisations (Fig.~\ref{figim}).

 We start from a turbulent initial condition (Fig.~\ref{figim} (a)). The first stage of the quench takes place for times $0\le t\lesssim 80$ and corresponds to a sudden decrease of both $e$ and $f$.
 During this stage, no laminar hole appears (Fig.~\ref{figim} (b)), however the turbulent structures increase in size and decrease in intensity.
 The most striking feature of this stage is that data from the five numerical experiments collapse on a master curve (Fig.~\ref{trempe3_bis}).
 This indicates an apparent deterministic behaviour,
 independent of the initial organisation. This feature can be used to make a study as function of $R_1$ of this stage: only one numerical experiment is necessary
 to capture the whole quantitative behaviour of the flow. The other interesting feature of this stage is that the behaviour of the flow does not depend
 on whether the system can accommodate bands or not. This suggests that this stage is dominated by a local process. Two types of behaviours appear:
 the average energy has an exponential-like decrease (Fig.~\ref{trempe3_bis} (a)), while the decrease of the turbulent fraction is linear (Fig.~\ref{trempe3_bis} (b)).
 This difference will be discussed in section~\ref{locm}.

  The second stage of the quench takes place for times $80 \lesssim t \lesssim 350$. It corresponds to the formation of the laminar troughs (Fig.~\ref{figim} (c)).
The organisation of the troughs is apparently random and bears no trace of the turbulent band yet. This stage can be divided into two parts: first,
both $e$ and $f$ undershoot. Then they grow back to the neighbourhood of their average value at this Reynolds number. Once they have reached this neighbourhood,
they fluctuate around it and do not vary significantly. Unlike the first stage of the quench, the evolution is more random and there is no correlation between each experiment.
The formation of laminar trough is discussed in more details in section~\ref{H}.

The last stage leaves no clear trace in $e$ and $f$ \cite{RM10_2}. This stage corresponds to the reorganisation of the holes
into the oblique laminar-turbulent bands (Fig.~\ref{figim} (d)). In all three stages, the rescaling of the velocity field by $R_0/R_1$ only causes small decreases of the time scales
and shifts of the minimum values of $e$ and $f$.

The duration of the last stage depends on the domain size $L_x\times L_z$ of the system and the complexity of the defects that
can appear in the band. Its duration is approximately of $1000h/U$ if the system accommodates one band. It is much longer if the system accommodates many bands:
There is a competition between domains of $+$ and $-$ orientations \cite{dsc10}. This case shares many similarities with coarsening dynamics \cite{bray}. If the comparison is quantitative,
one would expect scaling laws for the duration of the reorganisation stage as a function of the domain size $L_x\times L_z$.
The exponent depends on the type of system as well as its dimension.
The comparison is all the more tempting that Ginzburg--Landau equations, the prototype equation of coarsening, appears to be
a very good model of the measure of modulation $a_\pm(x,z,t)$ \cite{phD}.

\subsubsection{Effect of the arrival Reynolds number}

\begin{figure}
\centerline{\includegraphics[width=6.5cm]{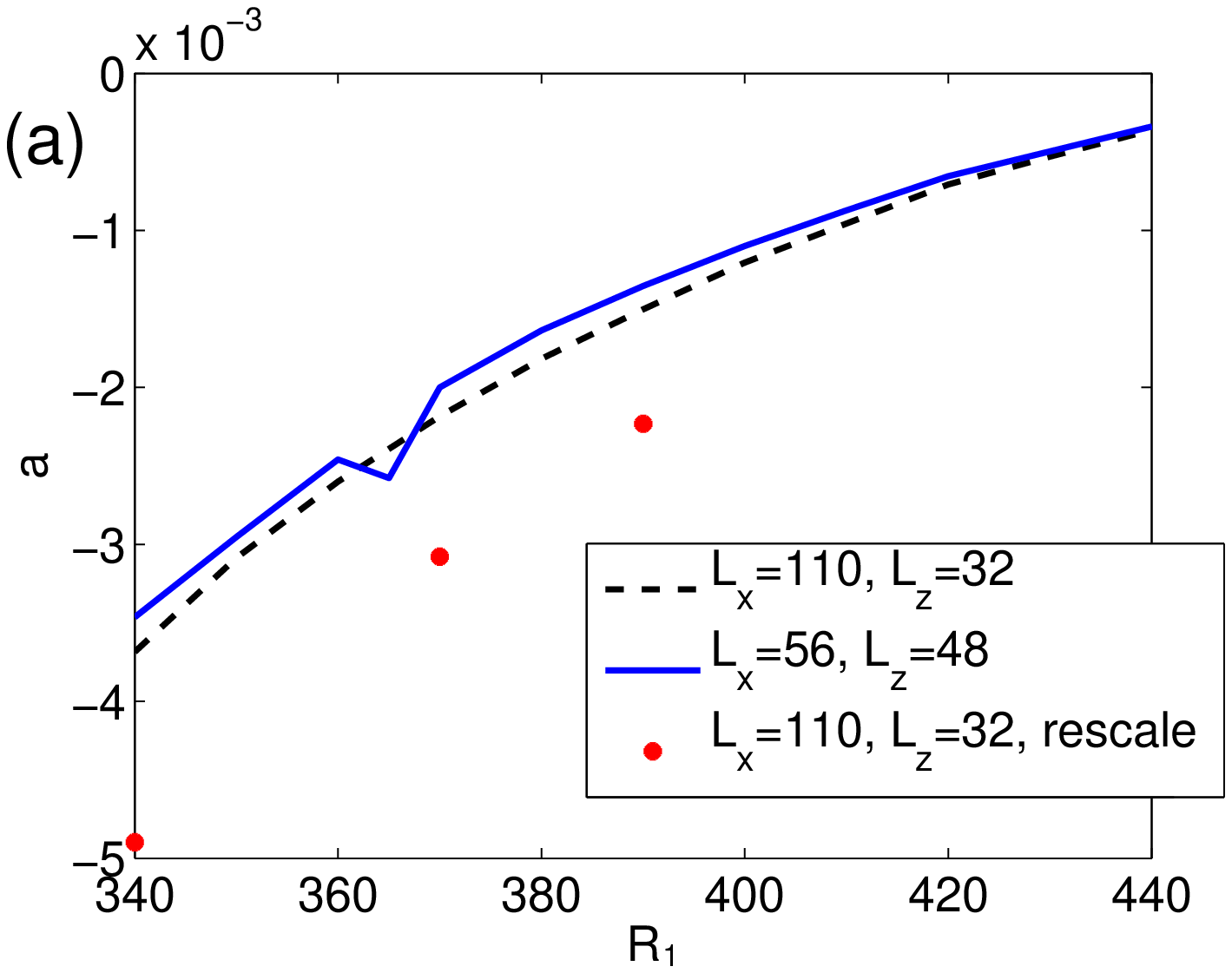}
\includegraphics[width=6.5cm]{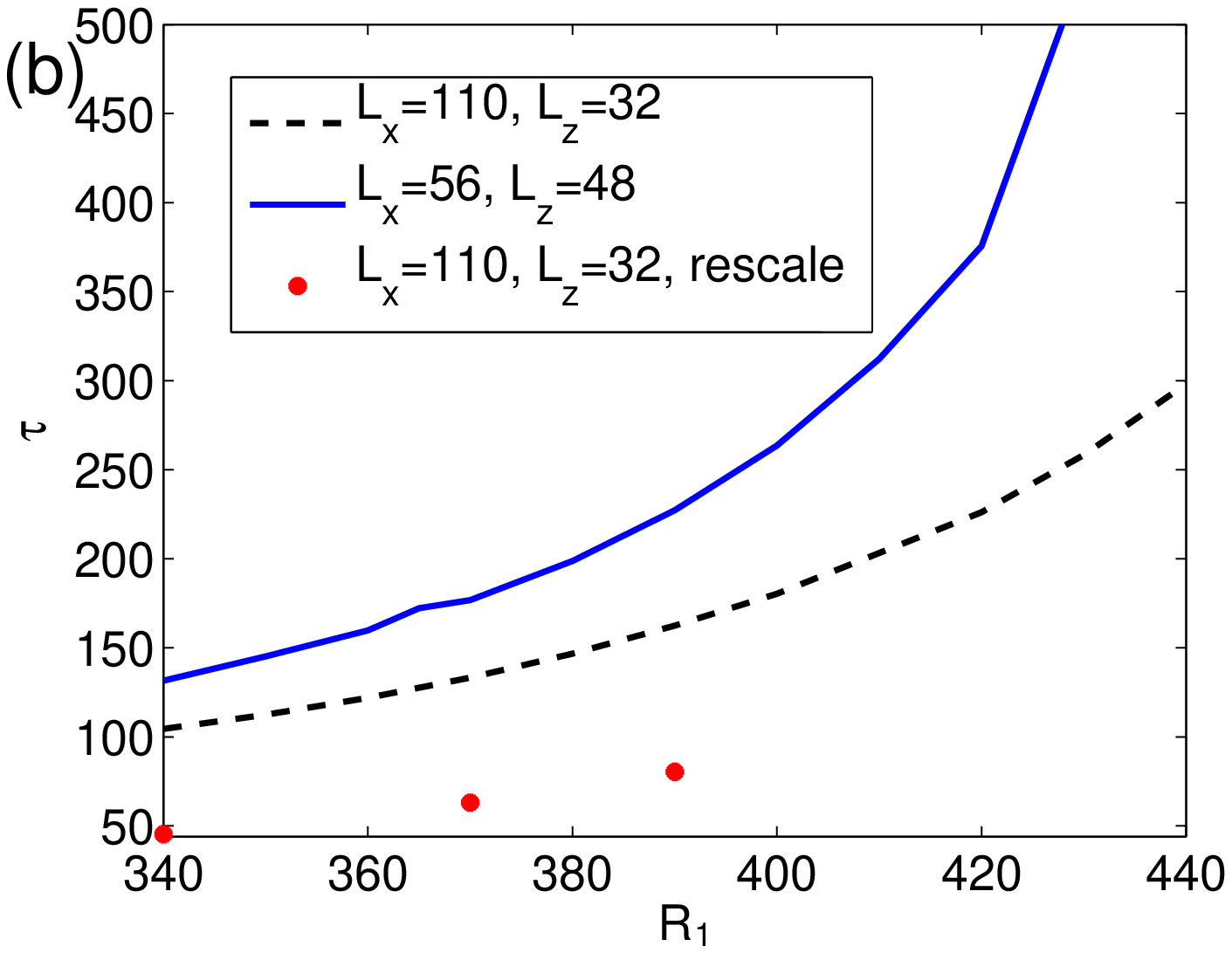}}
\caption{Fitted slope $a$ and growth rate $\tau$ of the time series of (a) : turbulent fraction,
(b) : average energy} \label{trempe2}
\end{figure}

A study of the quenches for a large range of arrival Reynolds numbers, $340\le R_1\le440$ is now performed for both sizes, using one numerical experiment per size and Reynolds number.
The same initial condition is used for each Reynolds number. Data sampled in quenches using a rescaled initial condition are added. Following the observation of the former section, the time series of the energy
are fitted by $\alpha \exp(-t/\tau)$ and those of the turbulent fraction by $at+b$.
The parameters $\tau$ and $-1/a$ correspond to the characteristic decay times of kinetic energy and the turbulent fraction respectively.
The parameters $\alpha$ and $b$ corresponds to the values of $e$ and
$f$ (respectively) in the featureless turbulence regime. One simply finds $b\lesssim 1$,
as is expected from the featureless turbulent regime. The slope of the turbulent fraction $a(R)$ increases toward zero with the Reynolds number (Fig.~\ref{trempe2} (a)).
It does not depend on the size of the domain in which the experiment is performed. The decay times $\tau(R)$ of the kinetic energy increases very regularly with $R_1$ (Fig.~\ref{trempe2} (b)).
A large incertitude appear at $R_1=365$, for $L_x\times L_z=56\times 48$: $a$ and $\tau$ have to be averaged over $5$ experiments. This can be caused by unusually short first
stages (Fig.~\ref{trempe3_bis} (a)).

The change of size is found to shift the values of $\tau$ by an additive constant. In both cases, for the lowest values of arrival Reynolds number $R_1$, the behaviour of $a$ and $\tau$ appears
to be independent of the value of the Reynolds number of the initial condition $R_0$.
However, $\tau$ and $-1/a$ diverge
as $R_1$ reaches $R_0=500$. Indeed, for $R_1$ close to $R_0=500$, the flow changes very little when the quench is performed, leading to approximately constant values of $e$ and $f$.
Both $-1/a$ and $\tau$ show that rescaling the velocity field as the Reynolds number is decreased leads to shorter time scales, but the same behaviour as a function of $R$.

\subsection{Local measurements \label{locm}}

This section investigates the quench at small scale. It aims at determining fate of a small part of the flow, given its initial condition.
For that matter, the domain is divided in small boxes of size $l_x\times l_z=6\times 4$ and thickness $l_y=1$ ($y>0$ or $y<0$). The size of the cell is chosen
so as to contain one or two velocity streaks. Around these values, no dependence of the result on the size was found. The state of the flow is evaluated
in these boxes by averaging the kinetic energy in the box: this gives a set of averaged kinetic energies $\{ e_{i}\}=\frac{1}{2}\int_{\text{box}} \parallel\textbf{v}\parallel^2\, {\rm d}x{\rm d}y{\rm d}z$.
Note that $\sum_i e_i=e$.
The study is performed in a domain of size of $L_x\times L_z=110\times 32$ with an arrival Reynolds number of $R_1=370$.
The set of $\{e_i(t)\}$ is studied as a function of $\{e_i(0)\}$, for $t$ ranging from $0$ to $250$.

\begin{figure}
\centerline{\includegraphics[width=6cm]{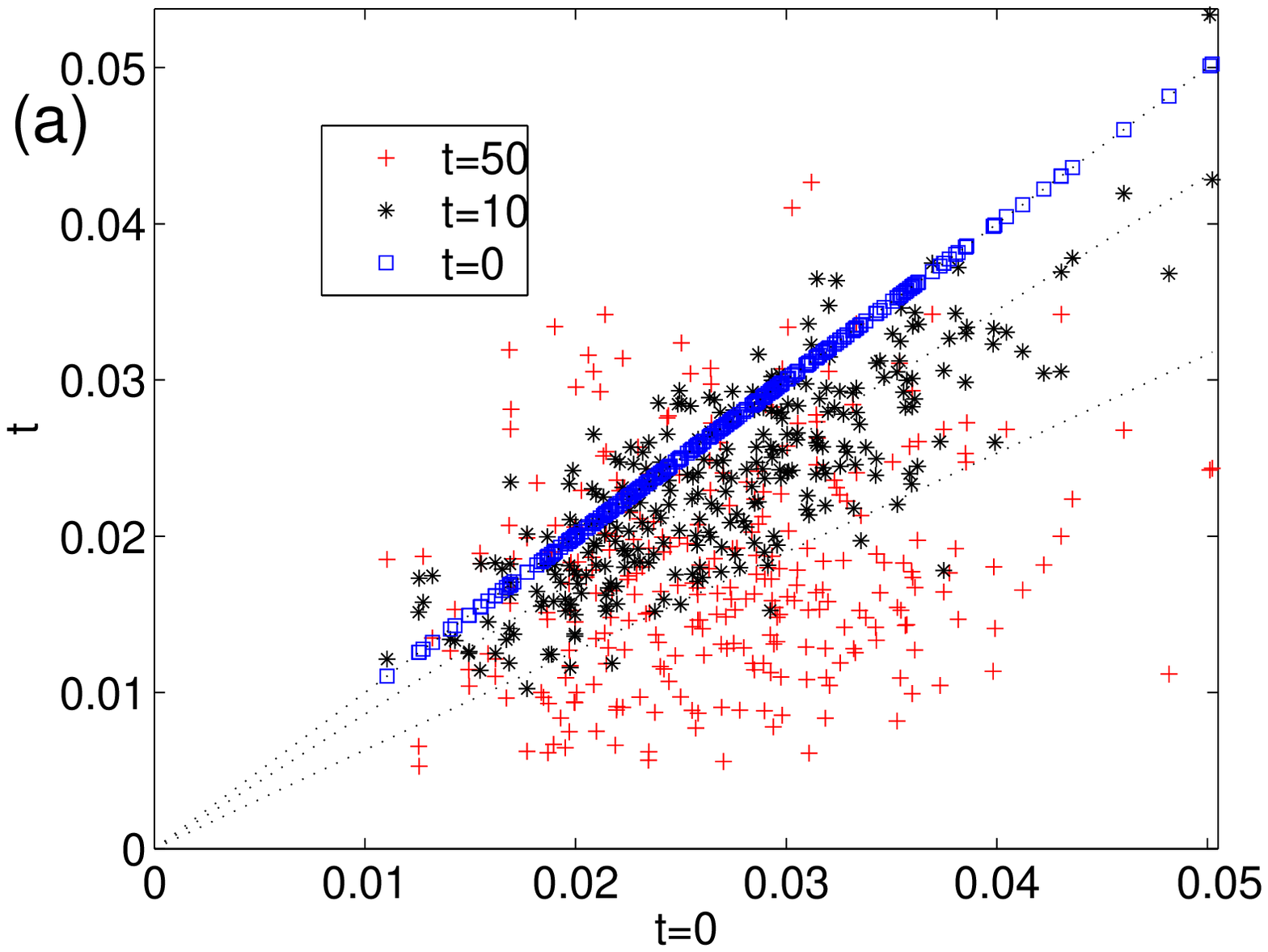}\includegraphics[width=6cm]{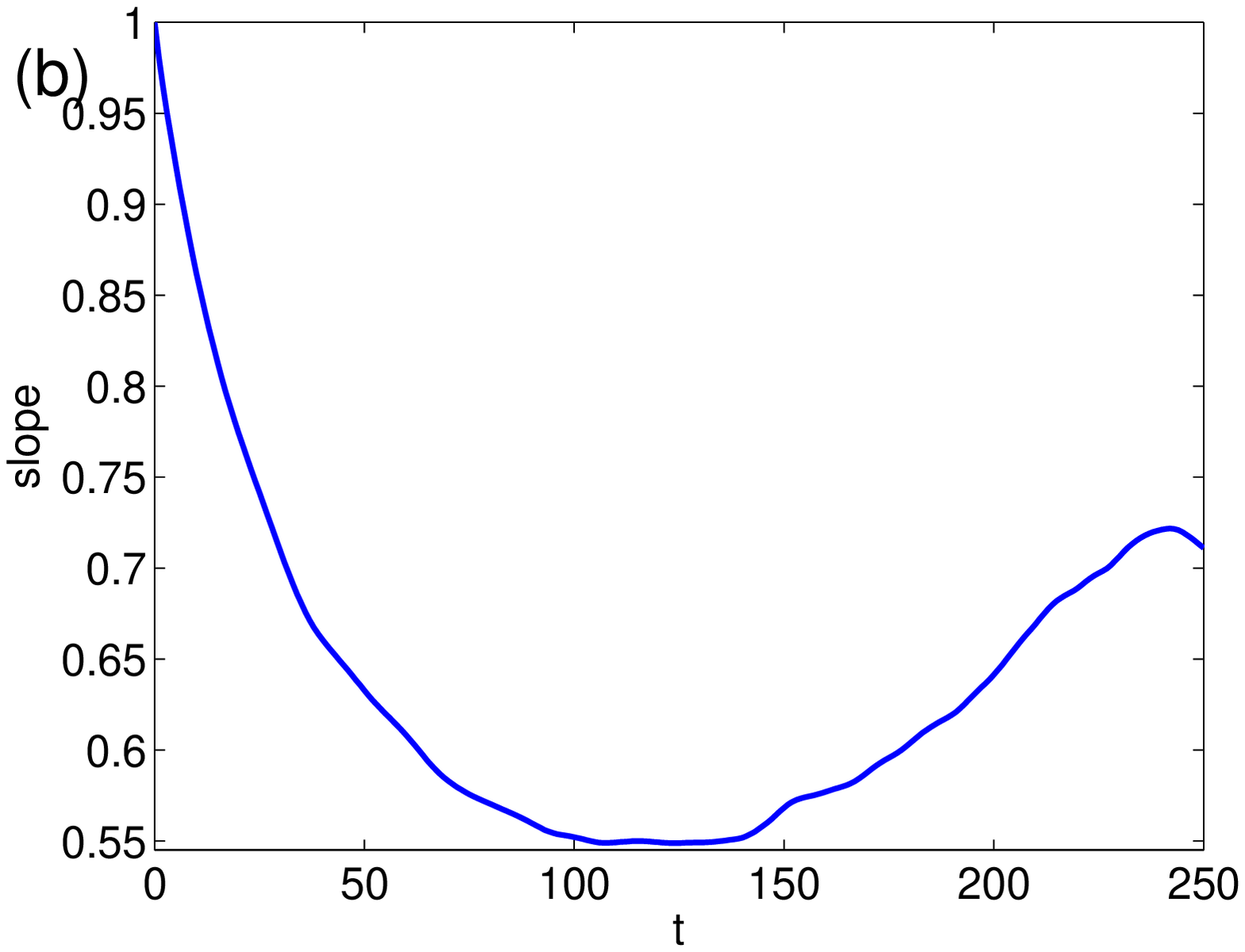}\includegraphics[width=6cm]{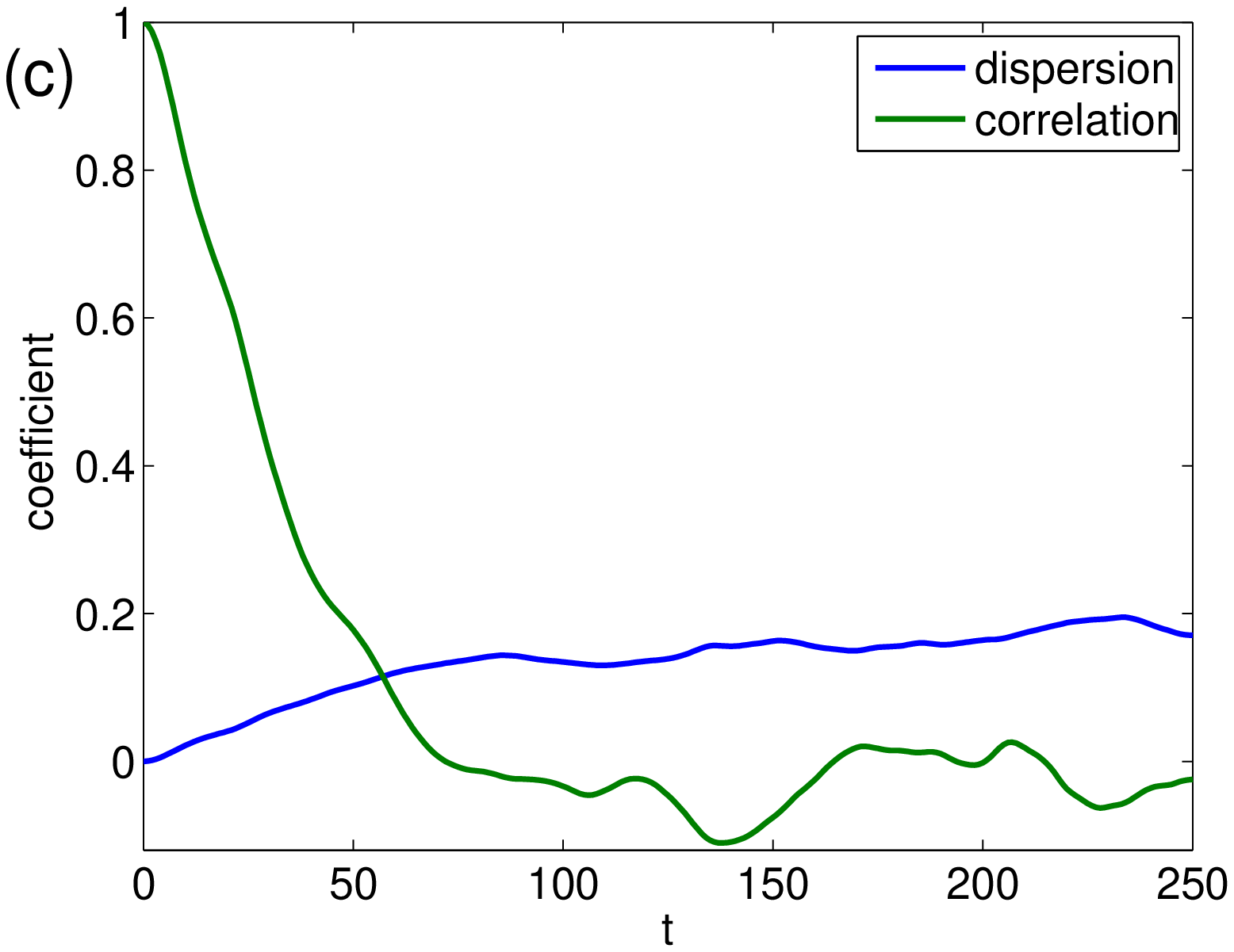}}
\caption{Local diagnostic of the quench $R_0=500\rightarrow R_1=370$, $L_x\times L_z=110\times 32$, $N_y=27$. Cloud of points :
(a) energy $\parallel\textbf{v}\parallel^2$ averaged in boxes at $t=0$, $t=10$, $t=50$ as a function of that at $t=0$. (b) : Slope of the clouds. (c) :
correlation and dispersion coefficients of the clouds as a function of time.} \label{trempe4}
\end{figure}

The result is presented as clouds of data and reveals two phenomena (Fig.~\ref{trempe4}(a)). The initial line becomes an expanding cloud,
and the cloud moves toward low values of $e$. The set of $\{e_{i} \}$ is not separated into two populations: one of smaller $e_i(0)$ that would
relaminarise and one of larger $e_{i}(0)$ that would stay turbulent. This confirms the observation of adjustment (Fig.~\ref{figim}). Note that during that stage, the distribution of $\{e_i\}$ is strongly positively skewed, with a relative skewness of approximately $+1$.

Several measures are used to quantify this observation. The slope of the cloud $a$ at time $t$ is defined by least squares:
\begin{equation}
\min_a\sum_i (e_i(t)-ae_i(0))^2 \Rightarrow a=\frac{\sum_i e_i(t)e_i(0)}{\sum_i e_i(0)^2}\,.\end{equation}
The normalised dispersion around the line is measured by the coefficients:
\begin{equation}
\text{dispersion}=\frac{\sum (e_i(t)-ae_i(0))^2}{\frac12\sum_i e_i(0)^2+e_i(t)^2} \,.
\end{equation}
And the correlation coefficient, \emph{i.e.} the sense of a linear fit of the data, by:
\begin{equation}
\text{correlation}=\frac{\sum (e_i(t)-e(t))(e_i(0)-e(0))}{\sqrt{\sum (e_i(t)-e(t))^2\sum(e_i(0)-e(0))^2}}
\end{equation}
The slope as a function of $t$ is displayed in figure~\ref{trempe4} (b). One can see the same exponential decay as that of the spatially averaged energy.
The characteristic decay time is approximately $100$, in agreement with data found at this Reynolds number (Fig.~\ref{trempe2}).
The cloud of $\{e_i\}$ disperses and decorrelates in a similar duration $t\simeq 50$ (Fig.~\ref{trempe4} (c)), as can be visualised in figure~\ref{trempe4} (a). This
 indicates that after this time,
the slope measures the behaviour of the average energy rather than a local decay of each cell.

Note that the two different behaviours of $f$ and $e$ in the first stage of the quench may be caused by the same phenomenon seen through different observables.
Indeed, the kinetic energy $e$ is governed by exponential decay at small and large scale. Meanwhile, the decrease of the turbulent fraction $f$ is more gradual.
It happens by small increments because the kinetic energy in a cell has suddenly decayed below the threshold $c$.

\section{Modeling the kinetic energy during the quenches \label{M}}

 The measurement of the kinetic energy:
\begin{equation}e=\frac{1}{2}\int\left((v_x)^2+(v_y)^2+(v_y)^2\right)\,{\rm d}x{\rm d}y{\rm d}z\end{equation}
 during the quenches can be compared to a budget derived from the Navier--Stokes equations.
Using periodic boundary conditions and tensorial summation conventions with indices $i,j,i'=\{1,2,3\}=\{x,y,z\}$, the budget reads \cite{schhu}:
\begin{equation}
\partial_t e=\underbrace{-\int v_x v_y\frac{dU}{dy}\,  {\rm d}x {\rm d}y {\rm d}z}_{\text{production}}
\underbrace{-\frac{1}{R}\int  (\partial_j
v_i)(\partial_j v_i)\,\prod_{i'}{\rm d}i'}_{\text{dissipation},<0}\,. \label{eqbilener1}
\end{equation}
This equation has one production term, energy extraction from the base flow, and a viscous dissipation term. In the case of dimensionless plane Couette
flow, the equation is simpler: $U(y)=y$ and $dU/dy=1$. When the flow is statistically steady,
one approximately has $\partial_t e\simeq 0$: the dissipation balances the production.

\begin{figure}
\centerline{\includegraphics[width=6cm]{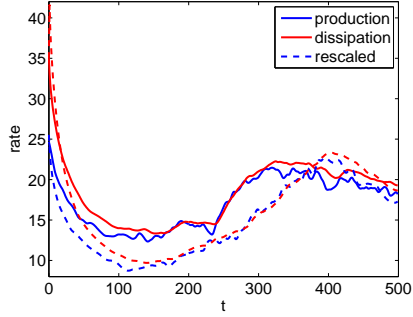}}
\caption{Production (blue lines) and dissipation rates (red lines) of the energy budget evaluated in a quench from $R_0=500$ to $R_1=370$ in a domain of size $L_x\times L_z=110\times 32$, for an augmentation of viscosity (full lines) and a reduction of velocity (dashed lines).}
\label{figbud}
\end{figure}

 Since the two terms of the right hand side have a structure different from the left hand side, this equation is not tractable as such. One can either estimate both terms from DNS (Fig.~\ref{figbud})
 or propose a simplified analytical treatment.

We can first examine the behaviour of both term \emph{via} DNS (Fig.~\ref{figbud}), using two decorrelated initial conditions,
with and without rescaling the velocity field by $R_{0}/R_{1}$. When the quench is performed, $1/R$, and therefore the dissipation,
is suddenly increased. As a consequence, one has $\partial_t e<0$. Both the dissipation and the production decay exponentially,
until they once again balance one another at time $t\simeq 150$. This corresponds to the end of the fast decrease of $e$,
and to the slower readjustment to the new Reynolds number. In this stage, $\partial_t e$ is much smaller (Fig.~\ref{trempe3_bis})
and leads mostly to fluctuations. One can see that the behaviour of the production and the dissipation is the same for both quench procedures:
in both cases, the exponential decay is driven by the sudden increase of dissipation between $t=0^-$ and $t=0^+$.

This behaviour can be described analytically. We first rewrite the budget. Indeed, the velocity $\mathbf{v}$ has periodic boundary conditions in $x,z$
and vanishing boundary conditions in $y$. This allows one to decompose $\mathbf{v}$ on a base of Fourier modes in $x,z$ and sine modes $\sin(((\pi n)/2) (y+1))$ in $y$,
introducing $k_y=\pi n/2$. Unlike Chebyshev polynomials, this choice is not adapted to the discretisation of the Navier--Stokes equations.
However, this provides a tractable way of writing the energy budget. Indeed, in the case of plane Couette flow, one finds:
\begin{align}\notag
e=\frac{1}{2}\sum_{k_x,k_z,k_y}\sum_i |\hat{v}_{i,k_x,k_y,k_z}|^2\,,\int  (\partial_j v_i)(\partial_j v_i)\,\prod_{i'}{\rm d}i'=  \sum_{k_x,k_y,k_z} \sum_i(k_x^2+k_y^2+k_z^2)|\hat{v}_{i,k_x,k_y,k_z}|^2\,,\,\\ \int v_x v_y\,{\rm d}x {\rm d}y {\rm d}z=\sum_{k_x,k_y,k_z}\hat{v}_{x,k_x,k_y,k_z}\hat{v}_{y,-k_x,k_y,-k_z} \,.
\end{align}
We make the hypotheses:
\begin{itemize}
\item The component $v_x$ dominates in the energy  as well as in the dissipation (see Fig.~\ref{figint}).
\item Instead of a wavepacket, the spectrum of $v_x$ is considered to be peaked around one given mode in $k_x$, $k_z$ and $k_y$, of wavelength corresponding
approximately to the coherence length of the streak $\lambda_x\simeq 20$, $\lambda_z\simeq 5$. Note that the assumption of a single mode is not much affected by non-linear effects:
indeed: the non-linearities conserve energy. The main drawback of this assumption is the possible dispersion of the wavepacket, caused mainly by the production term.
\end{itemize}
The energy is then $e=(1/2)|\hat{v}_x|^2(k_x,k_y,k_z)$, and the former equation reads:
\begin{equation}\frac{1}{2}\partial_t (|\hat{v}_x|^2)\simeq -|\hat{v}_y||\hat{v}_x|-\frac{k_x^2+k_z^2+k_y^2}{R}|\hat{v}_x|^2\,. \label{Eapprox}\end{equation}
Which can be rewritten and solved:
\begin{align}\notag \partial_t (|\hat{v}_x|)=-|\hat{v}_y|-\frac{k_x^2+k_z^2+k_y^2}{R}|\hat{v}_x|\, \Rightarrow \\ |\hat{v}_x|(t)=\underbrace{|\hat{v}_x|(0)\exp\left(-\frac{k_x^2+k_z^2+k_y^2}{R} t\right)}_{\text{decay}}+\underbrace{\int_{0}^t|\hat{v}_y|(t')\exp\left(-\frac{k_x^2+k_z^2+k_y^2}{R} (t-t')\right)\,{\rm d}t'}_{\text{adjustment to }v_y}\label{sol_bud} \end{align}
One finds exponential decay, caused by the initial unbalance between dissipation and production. This stops when extraction of energy from the base flow equals the viscosity. The adjustment to $v_y$ takes place over the same time scale.
In practice, $k_x^2$ is negligible compared to $k_z^2$ ($\lambda_z\simeq 4$) and $k_y^2$. This yields a decay time proportional to $R$:
 $\tau=(1/k^2)R$, with $1/k^2$ of order one for both $v_x$ and $e\simeq v_x^2$. This is consistent with the measure of $\tau$
 for the smaller values of $R$ (Fig.~\ref{trempe2} (b)). The budget is invariant under a rescaling of velocity by $R_1/R_0$,
 corresponding to a quench by decrease of velocity. The quality of the approximation (shape of wavepackets, time evolution of $v_y$) is slightly
 impacted by this change. This explains the shift found in $\tau (R)$ (Fig.~\ref{trempe2} (b)).

In the more realistic case of a time dependant profile $U(y,t)$ during the beginning of the quench, one has:
\begin{equation}\frac{1}{2}\partial_t (|\hat{v}_x|^2)\simeq -|\hat{v}_x|\int v_y\frac{dU}{dy}(t)\sin(k_y (y+1))\,{\rm d}y-\frac{k_x^2+k_z^2+k_y^2}{R}|\hat{v}_x|^2\,. \end{equation}
The same treatment of the budget can be done, $|\hat{v}|_y$ is replaced by $\int {\rm d}yv_y\frac{dU}{dy}(t)\sin(k_y (y+1))$ in equation~(\ref{sol_bud}).
This leads to the same time evolution: exponential decay due to viscosity until lift-up equals it. This shows that apparently artificial quenches are a good model of a laboratory experiment, in the sense that they are driven by the same physical mechanisms.

At a small scale, typically that of our averaging boxes, this description does not strictly hold. The boundary conditions are not periodic,
and flux terms that cancel out in average will appear in the budget:
\begin{align}\notag \partial_t e +\underbrace{\frac{1}{2}\int \,\sum_{j=x,z}\partial_j (v_j v_iv_i)\,{\rm d}x{\rm d}z+\frac{1}{2}\int \partial_x
(Uv_iv_i)\,\prod_{i'}{\rm d}i'}_{\text{momentum flux}}+
\underbrace{\sum_{j,j'=x,z}\int \, (1-\delta_{jj'})v_i\partial_{j'} v_i\,{\rm d}j{\rm d}2}_{\text{friction at boundaries}}+
\underbrace{\int  \partial_i (v_iP)\,\prod_{i'}{\rm d}i'}_{\text{incompressibility}}\\
=-\int v_xv_y\frac{dU}{dy}\,{\rm d}x{\rm d}y{\rm d}z-\frac{1}{R}\int (\partial_j
v_i)(\partial_j v_i)\,\prod_{i'}{\rm d}i'\,.\label{fullbil}\end{align}
 They correspond to advection of perturbation by the large scale flow, which usually excites turbulence \cite{prls,ispspot}. The integrals corresponds to the sum over the cells introduced in section~\ref{locm}.
They cause decorrelated and more random time evolution at small scale (Fig.~\ref{trempe4}).
However, the balance between viscous decay and extraction is expected to play the same role.
One can view equation~\ref{fullbil} as equation~\ref{eqbilener1} with an additional zero average noise term. This noise term explains why the distribution of $\{e_i \}$
remains around a definite trend even though the dispersion is relatively large (Fig.~\ref{trempe4} (a)).

\section{Holes in the bands \label{H}}

  We move from the formation of holes forced by the sudden change of Reynolds number to the natural apparition of
  holes inside the bands at a given Reynolds number. We present a case of change of orientation caused by the holes and reformation of bands.
  Then we propose a framework to study the fluctuations of quantities like the kinetic energy.

\subsection{Description}

\begin{figure}\centerline{\textbf{(a)}\includegraphics[width=7cm,clip]{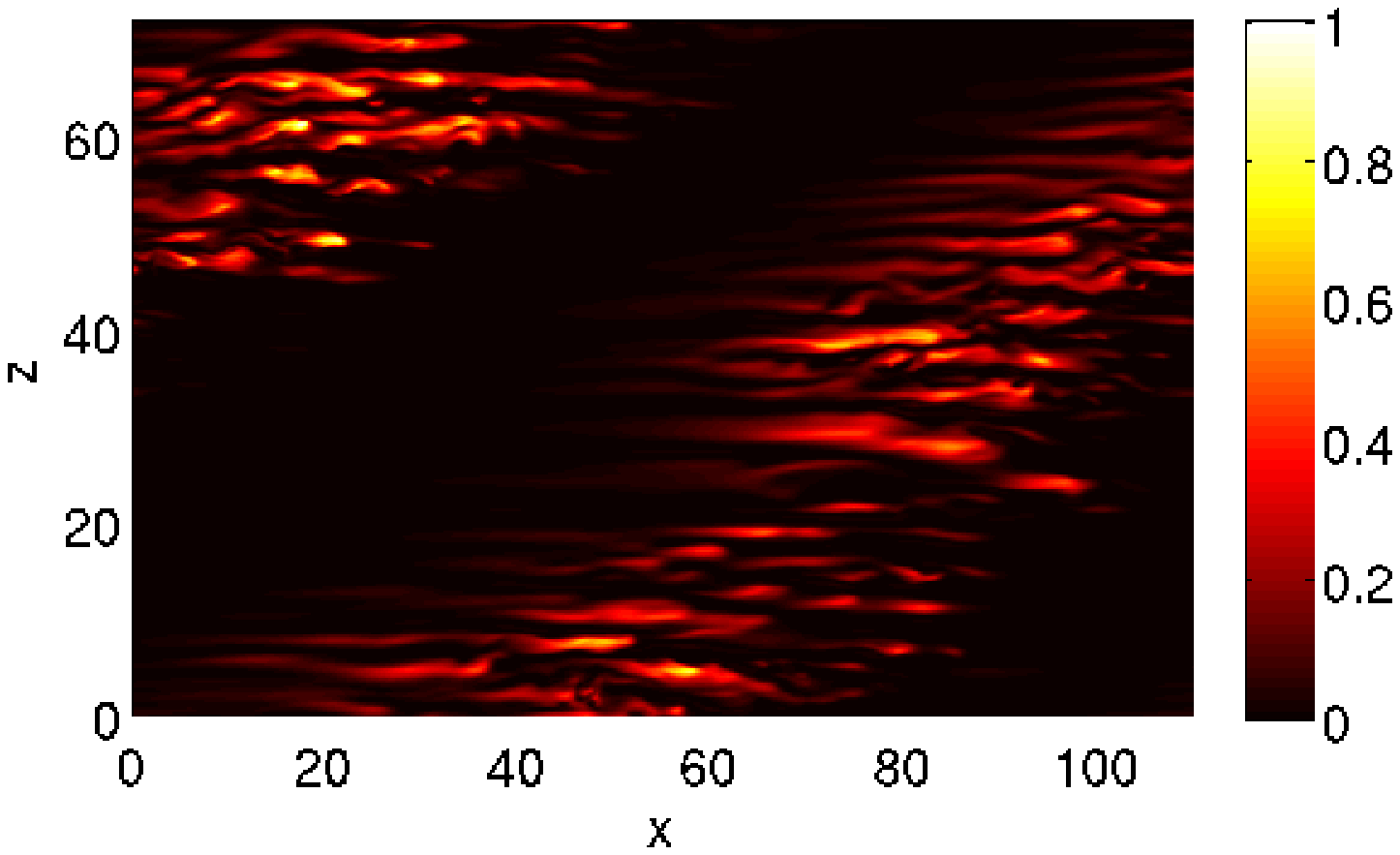}\textbf{(b)}
\includegraphics[width=7cm,clip]{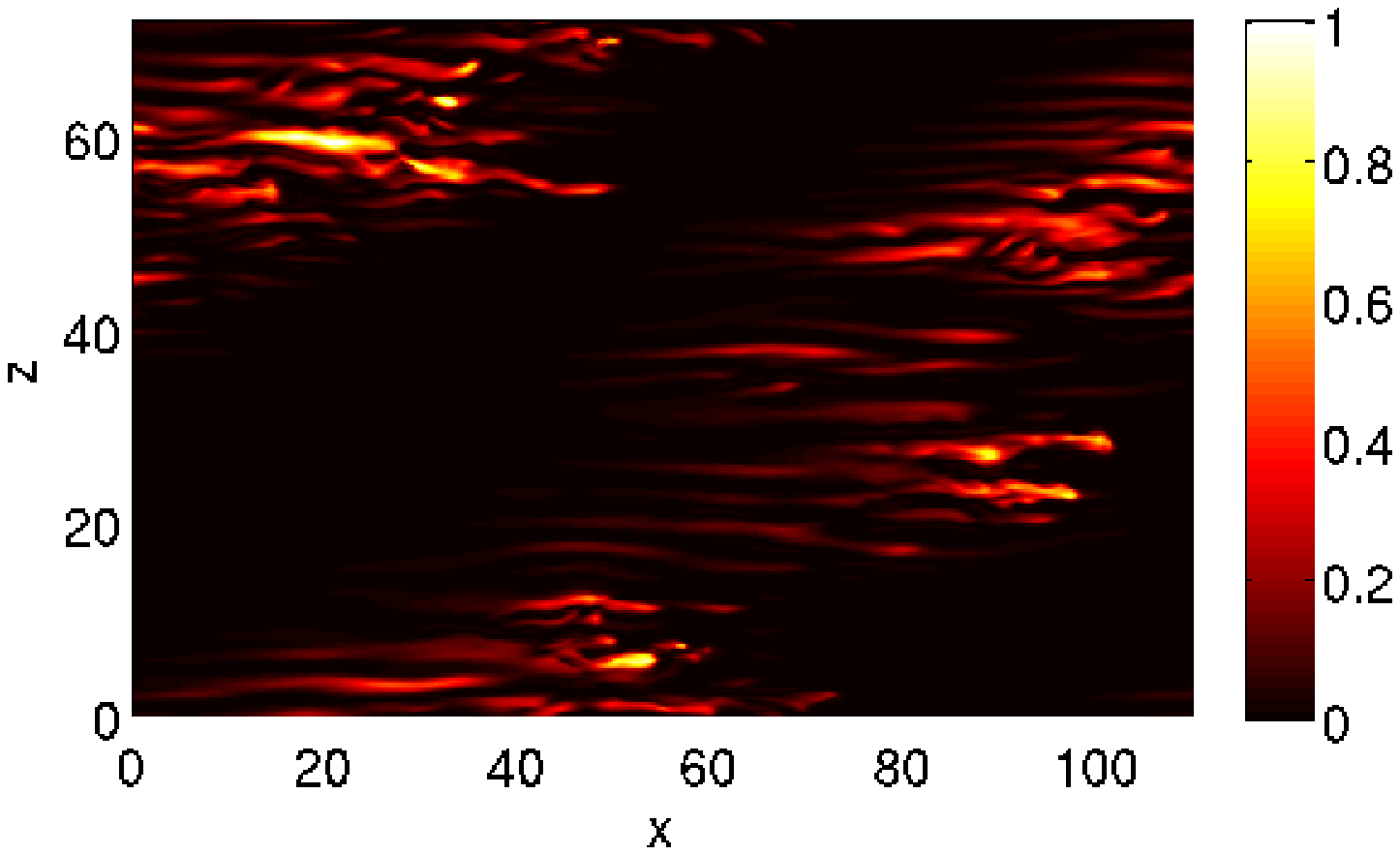}}
\centerline{\textbf{(c)}\includegraphics[width=7cm,clip]{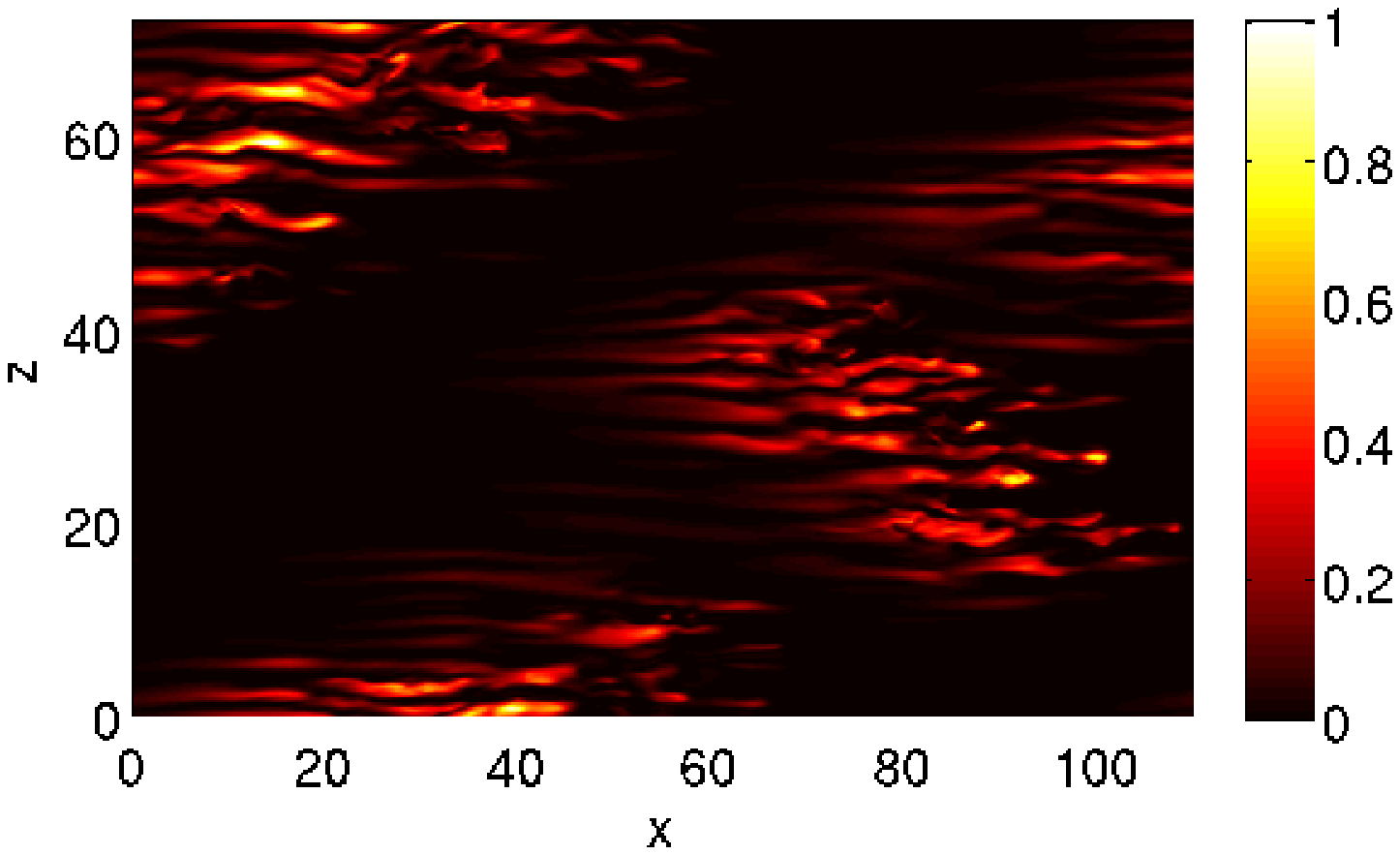}\textbf{(d)}
\includegraphics[width=7cm,clip]{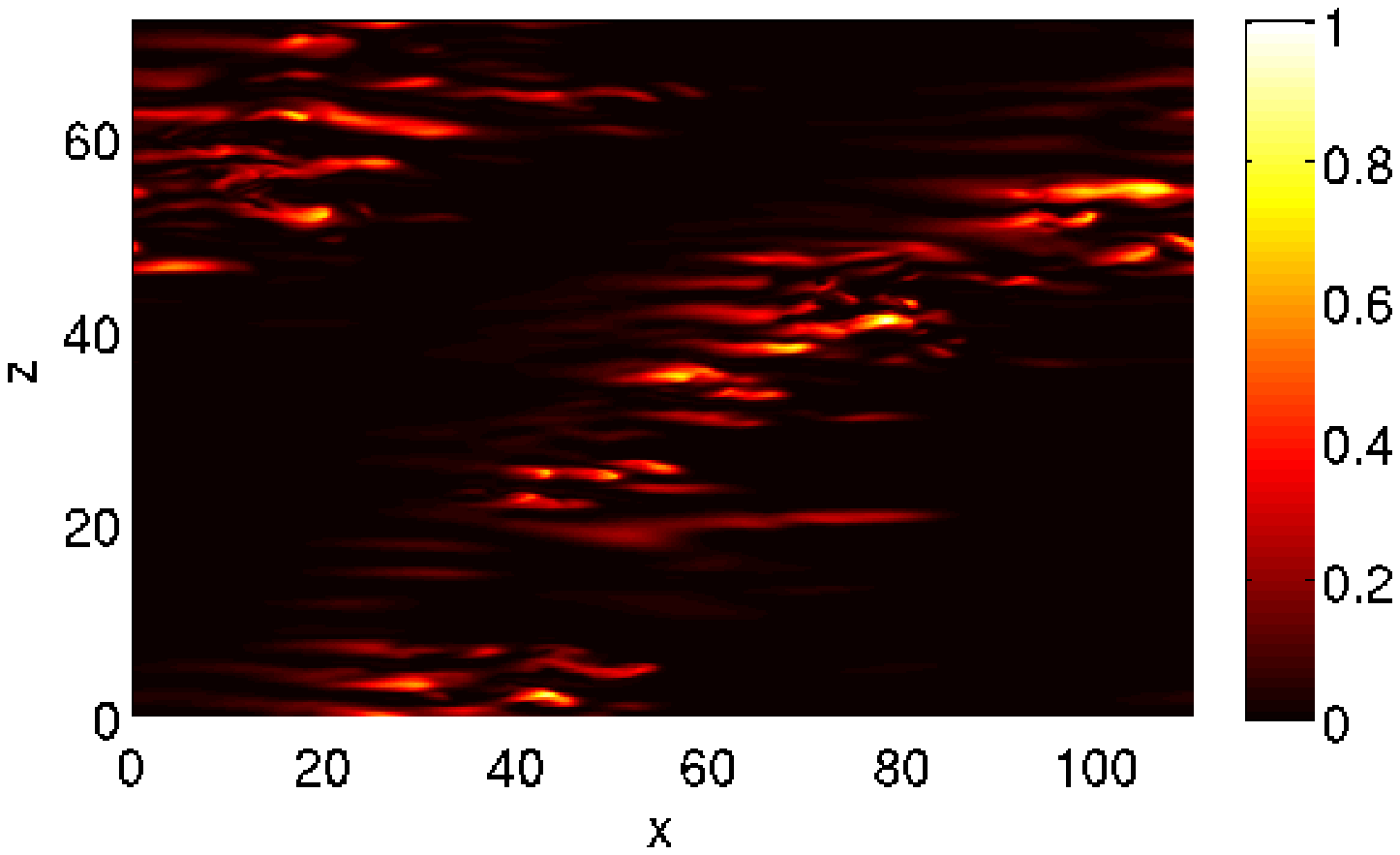}}
\caption{Norm of the departure to the laminar baseflow ${\vec v}^2$ in the
 $y=-0.62$ plane at four successive times in a domain of size $L_x\times L_z=110\times 72$, at $R=317$.. (a) : $T=3220$, (b) :
$T=3320$, (c) : $T=3450$. (d) : $T=3800$ (see timescale on figure~\ref{exc2}).} \label{exc1}
\end{figure}

\begin{figure}
\centerline{\includegraphics[width=6cm]{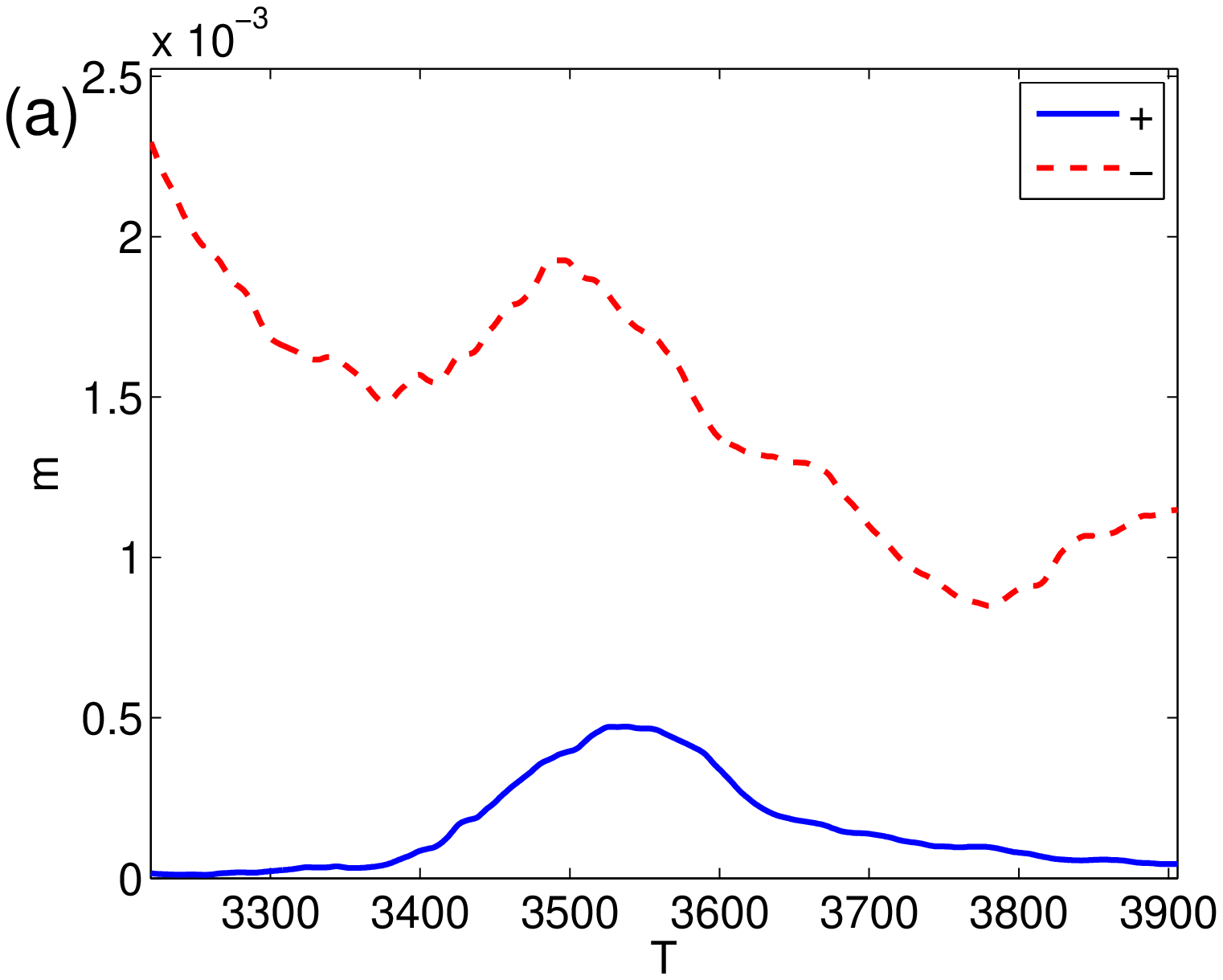}\includegraphics[width=6cm]{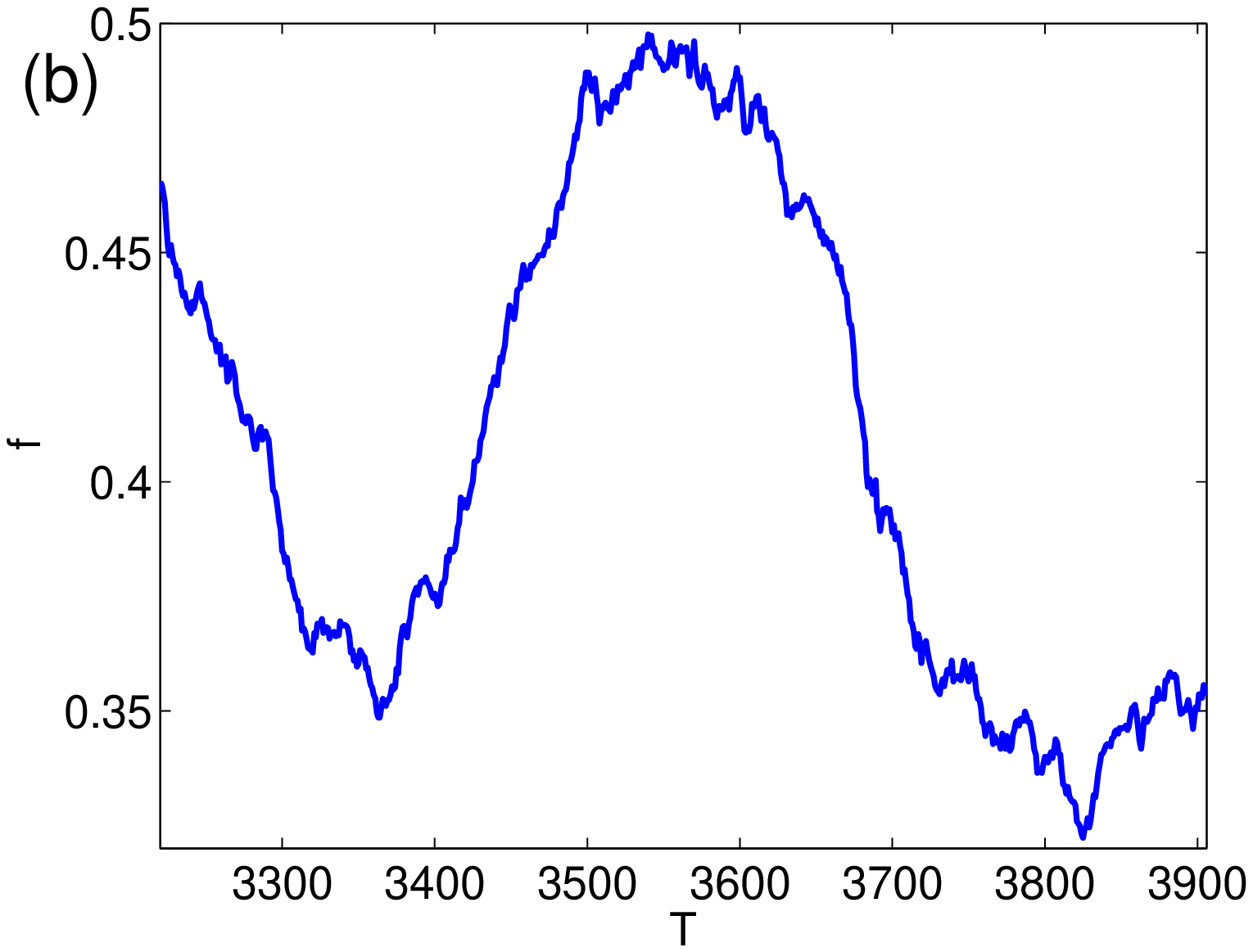}}
\caption{(a) : Time series of the order parameter and (b) of the turbulent fraction centered around the events of figure~\ref{exc1} ($L_x\times L_z=110\times 72$, $R=317$). } \label{exc2}
\end{figure}

  An example displaying hole formation and of change of orientation  at low Reynolds number is presented in figure~\ref{exc1}
(colour levels of $\parallel\textbf{v}\parallel^2$) and figure~\ref{exc2} (time series). The domain studied has a size of $L_x\times L_z=110\times 72$ (See Tab.~\ref{tab1}, $\sharp 3$).
The band is prepared at $R=330$, using a quench from $R=500$ and letting it adjust for a duration of $2000$.
The Reynolds number is slowly decreased toward $R=317$ in order to favour the formation of holes.

The domain contains one well formed wavelength of the band (Fig.~\ref{exc1} (a)). This corresponds to an order parameter dominating the other (Fig.~\ref{exc2} (a))
and a larger turbulent fraction (Fig.~\ref{exc2} (b)). After a long time ($\simeq 3000 h/U$), two holes appear at two different locations of the band (Fig.~\ref{exc1} (b)).
This event is concomitant  with a clear decrease of the turbulent fraction (Fig.~\ref{exc2} (b)) and a slight decrease of the dominating order parameter (Fig.~\ref{exc2} (a)).
A small germ of spot is left isolated in the flow, which grows instead of decaying. The growth takes an oblique, asymmetric band-like form (Fig.~\ref{exc1} (c)),
with an orientation opposite to that of the larger part of the band. This causes an increase of the turbulent fraction (Fig.~\ref{exc2} (b)),
and a decrease of the dominating order parameter while the dominated one grows (Fig.~\ref{exc2} (a)). Eventually, this spot hits the rest of the band,
which reorganises in the original orientation (Fig.~\ref{exc1} (d)), through the oblique advection of vortical structures by the large scale flow \cite{prls,ispspot}.

  These events can lead to changes of orientation near $R_{\rm g}$. They were seen in our low order modelling \cite{RM10_2} as well as in experiments \cite{phD}.
The same procedure of residency time measurements, based on the detection of the position of flow in phase space relatively to the two metastable states, can be used.
The relevant observable is the pair of order parameters $m_\pm$.
However, the physical modelling (Eyring--Kramers Theory of mean first passage times \cite{ha})
was specific and based on the fact that the system was well described by a gradient stochastic differential equation. The hole formation requires a more general approach.

\subsection{A general statistical approach ?}

  The formation of laminar holes, and more generally, the excursion of the turbulent flow toward the laminar state is a very rare event
for most Reynolds numbers \cite{shi,M11,io,gold}. The rarity of these events is simply quantified by their probability (or their probability per unit time, rate of probability).
From a physical point of view, these events usually correspond to a specific route followed by the flow \cite{M11}.

 Probability of rare events, long exponential tails, escape trajectories \emph{etc}. can be computed in the framework of Large Deviations,
a general approach of equilibrium and out of equilibrium statistical physics. It is centred around a Large Deviations principle (LDP). The LDP is simply
the convergence of the logarithm of the pdf $P$ toward a rate function $J$ as a parameter $\gamma$ of the system goes to zero:
\begin{equation}
J(x)=\lim_{\gamma\rightarrow 0}-\gamma\ln(P(x))\,.
\end{equation}
This is particularly interesting when the pdf is not a Gaussian. Indeed, when $\gamma$ is the inverse of the size or a number of particle,
it goes beyond results like the central limit theorem. This theorem only leads to Gaussian distributions and provides a quadratic approximation of $J$ around its minimum.

\begin{figure}
\centerline{\includegraphics[width=6cm]{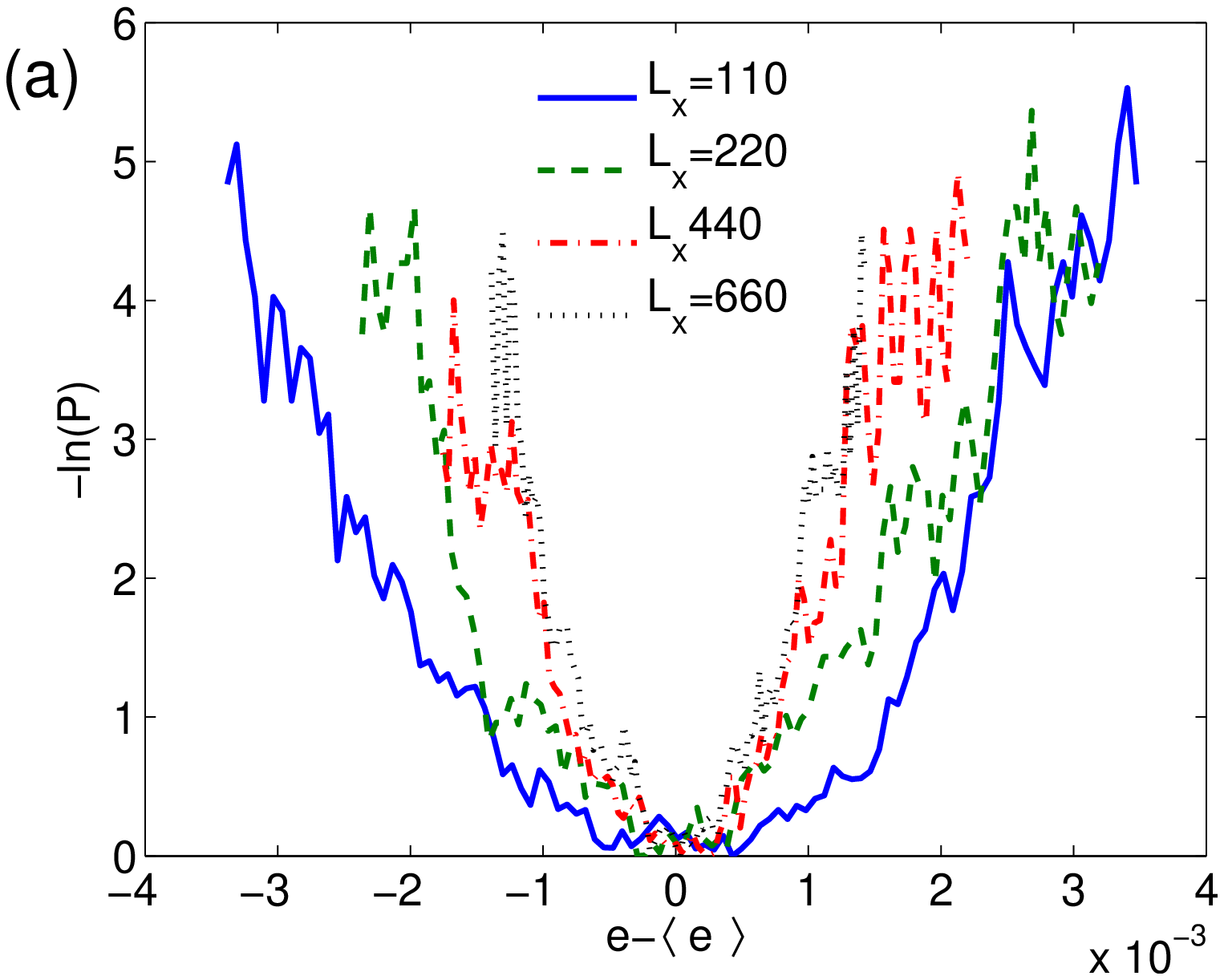}
\includegraphics[width=6cm]{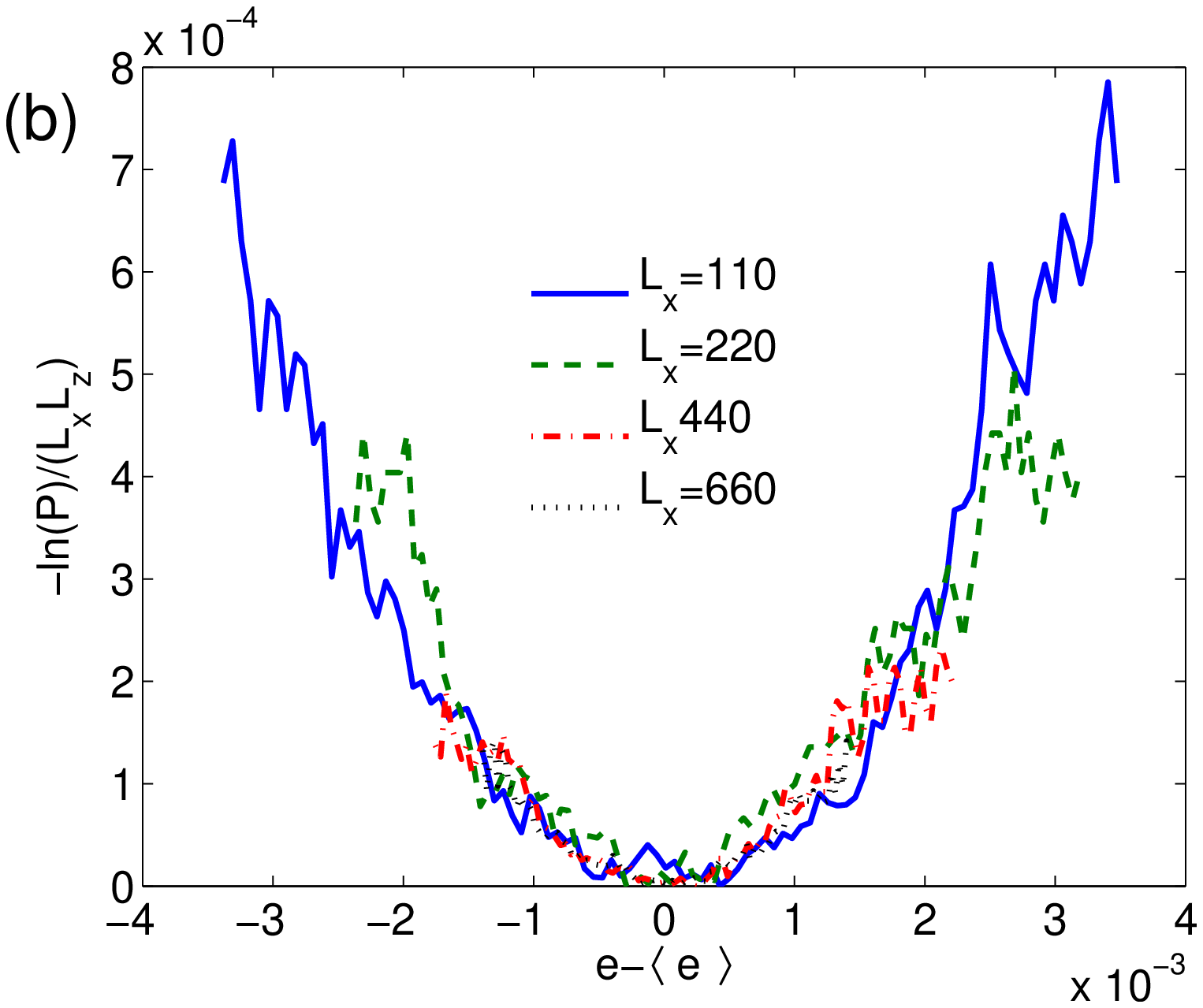}
\includegraphics[width=8cm]{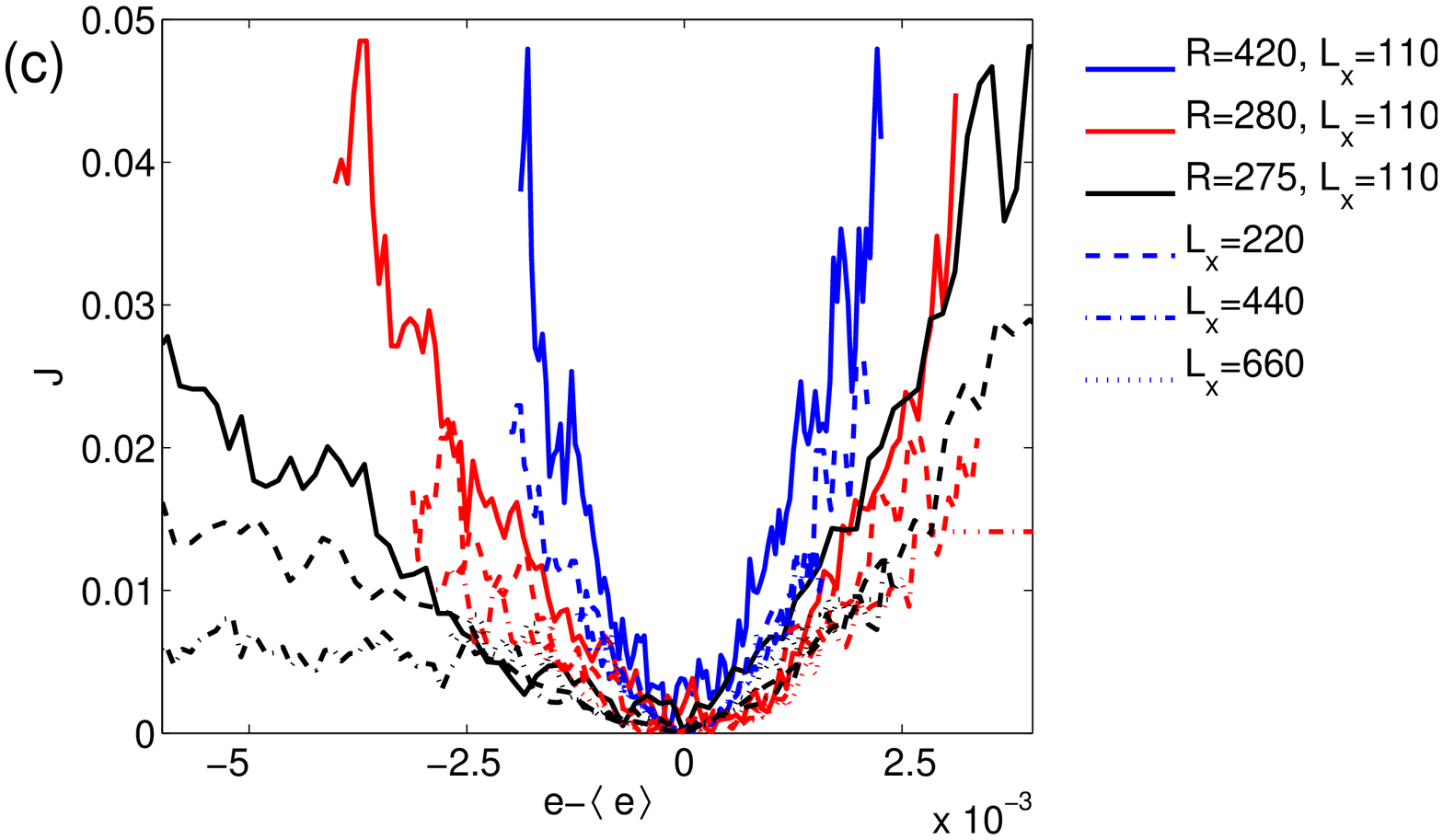}}
\caption{Processing of the pdf of the kinetic energy sampled with the low order modelling procedure ($N_y=15$), all functions are centred around the average of the energy. (a) : Logarithm of the pdf of the energy, at $R=310$, for four different sizes. (b) : Logarithm of the pdf of the energy rescaled by the size $L_x\times L_z$, at $R=310$ for four different sizes. (c) : Rate function $J$ of the energy, for three different Reynolds number, for the four different sizes.}
\label{rate}
\end{figure}

The kinetic energy is a good example to demonstrate that this framework can be applied to PCF. The formation of a hole in the bands,
which is the rare event, is obviously associated with an excursion of the average energy $e$ toward low values.
In PCF, the small parameter can be the inverse of the size of the system, and a rate function $J=\lim_{L_xL_z\rightarrow \infty}-\ln(P)/(L_xL_z)$.
In order to test that proposition, PDFs of $e$ in domains of increasing sizes, going from $L_x\times L_z=110\times 64$ to $L_x\times L_z=660 \times 48$ (Tab.~\ref{tab1}, $\sharp 4-7$),
are sampled with the low order procedure.

One can see that such a principle is well verified in the case of the pdf of the kinetic energy (Fig.~\ref{rate} (a,b),
where they are centred around $\langle e\rangle$). The fluctuations of the kinetic energy decrease as the size increases. There is also a
good collapse of $\ln(P)/(L_xL_z)$ on a master curve, even for the smallest size. This curve is identified with the rate function.
For a wide range of Reynolds number, the rate function is a parabola, which widens as the Reynolds number is decreased (Fig.~\ref{rate} (b,c)). This means that
fluctuations toward low or high values are equally probable. The rate function loses symmetry only for Reynolds number close to $R_{\rm g}$,
due to the changes in the phase space. This describes the exponential tail of the PDFs for small $e$, from which the probability of the formation
of a laminar hole can be extracted. In practice the loss of symmetry of the rate function shows to what extent the relaminarisation
becomes probable as $R$ is decreased toward $R_{\rm g}$.

\section{Discussion\label{D}}

This article presented a numerical study of the formation of laminar holes in low Reynolds number turbulence of plane Couette flow.
The hole formation was modelled using approaches from mechanics and statistical physics.

The first part (\S~\ref{Q}, \S~\ref{M}) was the use of so-called quenches, \emph{i.e.} sudden reductions of the Reynolds number from $R_0$ to $R_1$.
The DNS showed that the quantitative behaviour of the flow during the quench was rather independent of the way the quench was performed.
 Examination of the kinetic energy budget not only gave an explanation for the
decay rate of the kinetic energy, but also confirmed that all ways of performing the quenches were equivalent.

The second part (\S~\ref{H}) of this study was the natural formation of laminar holes into the turbulent bands near the Reynolds number
of disappearance of the bands. The holes appear when a local fluctuation is large enough to cross the basin boundary between
laminar and turbulent flow and can cause a local relaminarisation. In this case dissipation is too strong  to be balanced by the energy extraction  \cite{W}.
The holes can be closed by the advection of vortices by the large scale flow \cite{prls,ispspot}
or lead to a full relaminarisation of the flow \cite{M11}. However, our DNS showed an interesting alternative scenario:
if two laminar holes occur in the band at the same time, an isolated germ of band can grow with another orientation.
This is a mechanism leading to orientation fluctuation different from what is found near $R_{\rm t}$, in which the turbulent
fraction was nearly constant \cite{RM10_2}. This is very likely the mechanism responsible for the fragmented band regime observed in experiments \cite{phD}.

Besides, we verified that the extreme fluctuations toward low values of quantities like the kinetic energy could be
placed in the specific framework of Large Deviations when the size of the domain diverges. This was done by showing the asymptotic
convergence of the pdf in logarithmic form. This may provide general tools to compute average escape times, \emph{i.e.} the lifetime
of turbulence \cite{gold}, probabilities of relaminarisation. It may also provide approaches to determine paths between turbulent and
laminar flow that are different from the deterministic methods \cite{mnprl,PWK}. Indeed, they allow one to compute the full trajectory,
including the crossing of the separatrix. A first stage of this study could consist in studying the stochastic models of laminar-turbulent coexistence \cite{B}.
Indeed, most of the analytical large deviations results are worked out from stochastic or Markov chain models.

\section*{acknowledgements}
The author thanks F. Bouchet, Y. Duguet and P. Manneville for interesting discussions, the hospitality of the IAU of Frankfurt Goethe University,
where the modification were implemented and the comments from the anonymous referees that helped improve the manuscript.

\end{document}